\documentclass[12pt]{emulateapj}
\usepackage{natbib}

\newcommand{\ab}{_{\mbox{\small{AB}}}}
\newcommand{\smallab}{_{\mbox{\tiny{AB}}}}
\newcommand{\ubvi}{\hbox{$U\!BV\!I$}~}
\newcommand{\ubvriz}{\hbox{$U\!BV\!RI\!Z$}~}    

\slugcomment{The Astrophysical Journal Supplement Series, 162:20--37, 2006 January}

\shorttitle{Redshift Distributions in the CFDF--PRS}
\shortauthors{Brodwin et al.}

\notetoeditor{Please use a \figure* type environment (i.e. centered over the two columns) 
	for Figures 7,8,9,10 and 12}
\notetoeditor{Please use a \deluxetable* type environment (i.e. centered over the two columns) 
        for Tables 3,4,5,6,7 and 8 as shown in astro-ph/0310038}

\received{2003 October 2}
\begin{document}


\title{The Canada--France Deep Fields III:  Photometric Redshift Distribution to $I\ab = 24$}


\author{M.~Brodwin\altaffilmark{1}}
\affil{Department of Astronomy \& Astrophysics, University of Toronto, ON M5S 3H8, Canada}
\affil{Jet Propulsion Laboratory, California Institute of Technology, Mail Stop 169-506, Pasadena, CA 91109}

\author{S.~J.~Lilly\altaffilmark{1,2} and C. Porciani}
\affil{Department of Physics, Swiss Federal Institute of Technology (ETH--Z\"urich), ETH H\"oenggerberg, CH--8093, Z\"urich, Switzerland}

\author{H.~J.~McCracken}
\affil{Institut d'Astrophysique de Paris, 98 bis, bd Arago, 75014 Paris, France}

\author{O.~Le F\`evre\altaffilmark{1} and S.~Foucaud}
\affil{Laboratoire d'Astrophysique de Marseille, Traverse du Siphon, 13376 Marseille Cedex 12, France}

\author{D.~Crampton}
\affil{Herzberg Institute of Astrophysics, West Saanich Road, Victoria, BC V9E 2E7, Canada}

\author{Y.~Mellier}
\affil{Institut d'Astrophysique de Paris, 98 bis, bd Arago, 75014 Paris, France}


\altaffiltext{1}{Visiting Astronomer, Canada--France--Hawaii Telescope, 
operated by the National Research Council of Canada, the Conseil
nationale de recherche scientifique and the University of Hawaii.}
\altaffiltext{2}{Visiting Astronomer,  Cerro Tololo Inter American 
Observatory and Mayall 4--meter Telescopes, divisions of the National
Optical Astronomy Observatories, which are operated by the Association
of Universities for Research in Astronomy, Inc. under cooperative
agreement with the National Science Foundation.}


\begin{abstract}
  
  We compute accurate redshift distributions to $I\ab = 24$ and $R\ab
  = 24.5$ using photometric redshifts estimated from six--band \ubvriz
  photometry in the Canada--France Deep Fields--Photometric Redshift
  Survey (CFDF--PRS).  Our photometric redshift algorithm is
  calibrated using hundreds of CFRS spectroscopic redshifts in the
  same fields.  The dispersion in redshift is $\sigma/(1+z) \la 0.04$
  to the CFRS depth of $I\ab = 22.5$, rising to $\sigma/(1+z) \la
  0.06$ at our nominal magnitude and redshift limits of $I\ab = 24$
  and $z \le 1.3$, respectively.  We describe a new method to compute
  $N(z)$ that incorporates the full redshift likelihood functions in a
  Bayesian iterative analysis and we demonstrate in extensive Monte
  Carlo simulations that it is superior to distributions calculated
  using simple maximum likelihood redshifts.  The field--to--field
  differences in the redshift distributions, while not unexpected
  theoretically, are substantial even on
  30\arcmin\ scales.  We provide $I\ab$ and $R\ab$ redshift
  distributions, median redshifts, and parametrized fits of our
  results in various magnitude ranges, accounting for both random and
  systematic errors in the analysis.

\end{abstract}



\keywords{Galaxies: Distances and Redshifts---Galaxies: Photometry---Methods: Statistical---Methods: Data Analysis}


\section{Introduction}

The last decade has seen great advances in our understanding of the
evolution of the Universe, primarily due to groundbreaking studies of
the intermediate \citep{cfrs1} and high redshift \citep{steidel96}
Universe.  These "pencil beam" surveys probed to cosmological depths
over quite modest areas, revealing the strong luminosity,
morphological, and clustering evolution of the galaxy population over
the last 12 Gyr.  It was, however, recognized at the time that the
small windows on the Universe provided by these surveys were subject
to substantial sample, or ``cosmic'' variance, in particular for
measurements on the scale of the survey field sizes.

Recently the 2dF \citep{2df} and SDSS \citep{sdss} surveys provided
the first precise "local" ($z \sim 0.1$) measurements of the galaxy
luminosity function (\citealt{norberg_lf}; \citealt{sdss_lf}),
providing a crucial baseline reference for evolutionary studies.  In
addition, measurements of the galaxy correlation function in these
surveys \citep{norberg_cf,sdss_cf} have convincingly demonstrated
luminosity--dependent clustering long seen at lower significance in
smaller samples.  These surveys finally overcame the cosmic variance
issues that had plagued previous local surveys by sampling over 10$^3$
square degrees.

With the advent of large format mosaic CCDs, we undertook the
Canada--France Deep Fields (CFDF) survey, a deep \ubvi imaging survey
in four $30\arcmin \times 30\arcmin$ fields.  Published results from
the CFDF include a study of the galaxy angular correlation function to
$I\ab=25$ \citep{cfdf1}, and a measurement of the clustering
properties of Lyman--break galaxies at $z\sim 3$ \citep{cfdf2}.  In
this paper we introduce the CFDF Photometric Redshift Survey
(CFDF--PRS), derived from a highly uniform \ubvriz sub--sample of the
main survey, which forms the basis of several evolutionary
measurements of the galaxy population as a function of redshift.
Future papers (Brodwin et al, in preparation) will present studies of
the galaxy luminosity and correlation functions.

In this paper we focus on an accurate measurement of the galaxy
redshift distribution from $0 < z < 1.3$.  We introduce a novel
iterative technique to extract the optimal redshift distribution using
the full redshift likelihood functions for each galaxy.  We present
binned redshift distributions along with parametrized fits to our
results.

The outline of this paper is as follows.  Section
\textsection{\ref{section cfdf}} briefly describes the CFDF--PRS data.
Section \textsection{\ref{section photoz}} describes the calculation
and calibration of the photometric redshifts.  Section
\textsection{\ref{iterated method}} introduces the iterative Bayesian
method of recovering the redshift distribution using the full redshift
probability function for each galaxy.  In Sections
\textsection{\ref{section: zdist}} and \textsection{\ref{Section:
    fits}} we present the photometric redshift distributions, obtained
using the iterative method, as a function of sample limiting magnitude
in the CFDF down to $I\ab = 24$ and $R\ab = 24.5$, both in half
magnitude bins and via parametrized fits.  Finally, in Section
\textsection{\ref{summary}} we summarize our results.

Most of our results are independent of cosmology.  Where necessary we
have assumed a concordance cosmology in agreement with the recent WMAP
\citep{wmap} results $\{\Omega_M,\Omega_\Lambda\} = \{0.27, 0.73\}$
and a Hubble constant $h=0.71$ where $H_0 = 100\,h$.  All distances
are expressed in comoving Mpc.

\section{Canada--France Deep Fields}
\label{section cfdf}
The Canada--France Deep Fields (CFDF) is a deep, multi--color imaging
survey covering 1 deg$^2$ (\citealt{cfdf1}; \citealt{cfdf2}, hereafter
CFDF1 and CFDF2, respectively).  The acquisition, reduction, and
verification of the astrometric and photometric integrity of the
original \ubvi data are described extensively in CFDF1.  This first
paper in the series also includes extensive completeness simulations
showing that the survey is $\sim 90\%$ complete to $I\ab \leq 24$.
Deep $B\ab$ and $I\ab$ galaxy counts along with a detailed angular
clustering analysis confirm the reliability of the data to at least
this depth.

\subsection{Supplementary $\mathbf{R}$ and $\mathbf{Z}$ Imaging}

Photometric redshift simulations described elsewhere \citep{brodwin99}
indicated the need for additional $R$-- and $Z$--band imaging to
produce accurate photometric redshifts out to $z \sim 1.3$.  Further
deep CFHT imaging was obtained in these bands, along with some
additional $I$ and $B$ imaging to replace time originally lost to bad
weather.  We replaced the UH8k data with this superior, deeper CFH12k
$I$ and $B$ data, since combining the two datasets would have degraded
the excellent CFH12k seeing and cosmetic quality and while adding
little in terms of depth. These data were reduced, stacked and
calibrated using standard procedures (e.g.~bias subtraction, flat
fielding, fringe removal, geometric distortion correction), carefully
adapted to mosaic CCD imaging as described in CFDF1 and
\citet{thesis}.

New catalogues for this work were generated using the $\chi^2$
technique of \citet{szalay_chi2}.  The various magnitude limited
galaxy samples studied in this paper were selected using total
magnitudes as measured by the Sextractor \citep{sextractor} MAG\_AUTO
parameter, whereas the photometric redshifts are computed using
2.5\arcsec\ diameter aperture magnitudes.

Star/galaxy classification was performed using a hybrid technique that
incorporates both morphology (in 0.6\arcsec $Z$--band images) and 4
color-color planes of information (see \citealt{thesis} for full
details).  Only objects extremely reliably classified as stars are
removed from the present analysis.

\subsection{Photometric Redshift Sample}

As explained in CFDF2, $U$--band imaging only exists in 2.5 of our 4
$30 \arcmin \times 30 \arcmin$ fields.  In addition, data in one of
the eight UH8k CCDs with which the original data was taken is not
science grade, due to bad charge transfer efficiency.  Limiting the
survey to those regions containing 6--band photometry, and excluding
the regions masked due to bright (and saturated) stars, scattered
light, remnant cosmetic defects that survived the median stacking, and
the aforementioned unreliable CCD, resulted in a final
CFDF--Photometric Redshift Survey (CFDF--PRS) area of 0.45 deg$^2$.
Although such conservative cuts remove over half of the original
survey area from the PRS, the remaining data is of very high quality
and uniformity, and has the wavelength coverage necessary for robust
photometric redshifts.  The final photometric sensitivities for the
CFDF--PRS are presented in Table \ref{photometry}.

\begin{deluxetable}{ccccccc}
\tabletypesize{\smallsize}
\tablecaption{Final Photometric Sensitivities in the CFDF--PRS.  \label{photometry}}
\tablewidth{0pt}
\tablehead{
\colhead{Field} & \colhead{R.A.} & \colhead{Dec.}
  &  \colhead{Area} & \colhead{Band}  & \colhead{5$\sigma$/2.5\arcsec} &  \colhead{Seeing} \\
 & \colhead{(2000)} & \colhead{(2000)}
  &  \colhead{(deg$^2$)} & \colhead{}  & \colhead{(AB mags)} &  \colhead{(arcsec)}
}
\startdata \\ 
03hr           &  03:03:00 & +00:09:00 & 0.17 & U & 25.17 & 1.0\\
           &                    &  &  & B & 25.41 & 0.80\\
           &                    &  &  & V & 25.04 & 1.3\\
           &                    &  &  & R & 25.15 & 0.95\\
           &                    &  &  & I & 25.04 & 0.70\\
           &                    &  &  & Z & 24.26 & 0.65\\ \\ \tableline\\
14hr           &  14:17:10 & +52:24:43 & 0.18 & U &25.88 & 1.4\\
           &                    &  &  & B & 25.33 & 0.80\\
           &                    &  &  & V & 25.04 & 1.0\\
           &                    &  &  & R & 25.21 & 0.95\\
           &                    &  &  & I & 24.51 & 0.70\\
           &                    &  &  & Z & 24.47 & 0.70\\ \\ \tableline\\
22hr           &  22:17:48 & +00:17:13 & 0.10 & U &25.96 & 1.4\\
           &                    &  &  & B & 25.23 & 0.8\\
           &                    &  &  & V & 25.18 & 1.0\\
           &                    &  &  & R & 25.42 & 0.90\\
           &                    &  &  & I & 25.36 & 0.75\\
           &                    &  &  & Z & 24.65 & 0.52\\
\enddata
\end{deluxetable}

\section{Photometric Redshift Algorithm in the CFDF--PRS}
\label{section photoz}
In order to estimate photometric redshifts and calculate the redshift
probability distributions needed for the present study as well as the
luminosity and clustering analyses of future CFDF papers (Brodwin et
al.\ in preparation), we employ a standard empirical template fitting
algorithm, in which filter-convolved empirical galaxy template SEDs
are fitted to the observed galaxy photometry to determine the
best--fitting redshift and spectral type.  To obtain robust galaxy
colors the photometry in each filter is co--convolved to the worst
seeing in each field (1.3\arcsec-1.4\arcsec).  The photometry is
fitted to the templates in the flux domain, with the flux error for
each object taken as the quadrature sum of its Poisson noise and a
Monte Carlo estimate of the sky error in the aperture.

The \citet[hereafter CWW]{cww} empirical SEDs, supplemented by the
\citet{kinney96} SB2 and SB3 starburst SEDs, were adopted as the basis
templates.  This combination of templates has been shown to span the
range of galaxy properties in the HDF from $0 < z < 6$
\citep{benitez,fernandez-soto99}.  The CWW templates were extended into
the far UV using normalized solar metallicity GISSEL models.  The
\citet{kinney96} starburst SEDs, well measured down to $\sim 1100$~\AA,
were extended to 912~\AA\ via simple linear extrapolation.  Fifteen
interpolates of these basis spectra were used to improve the redshift
resolution.  Galaxy reddening is naturally incorporated into our
empirical templates, as they are each composite spectra from numerous
galaxies with varying levels of extinction.  Our SEDs are
statistically corrected for intergalactic hydrogen absorption as a
function of redshift following \citet{madau95}, taking into account
both Lyman series line blanketing at wavelengths below
$\lambda_{\mbox{\scriptsize obs}} < 1216\,(1+z)$ \AA\ and continuum
photoelectric absorption below the Lyman limit,
$\lambda_{\mbox{\scriptsize obs}} < 912 \,(1+z)$ \AA.  A maximum
likelihood method (ML) was used to produce the redshift likelihood
surface for each object as a function of redshift and spectral type.

\begin{figure}[hbtp]
\epsscale{1.25}
\hspace*{-0.6cm}
\plotone{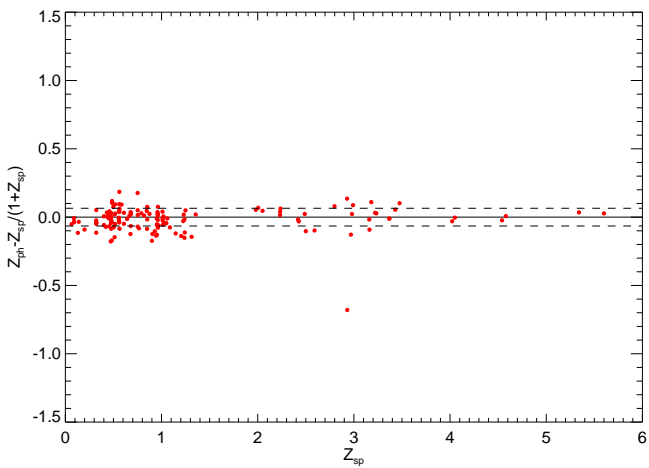}
\caption{Photometric redshift residuals for spectroscopic galaxies in the
HDF North for the sample defined in \citet{fernandez-soto02}.  This
Figure is directly comparable to their Figure 7.  The dashed lines
enclose the 1$\sigma$ error interval, which, excluding the single
discrepant galaxy, is $\Delta z/(1+z) = 0.07$.}
\label{Fig: hdf}
\end{figure}

This code was tested extensively with public HDF data and using the
hundreds of CFRS spectroscopic redshifts in our fields.  The latter
sample allowed a novel form of calibration of the method, described
below, to significantly improve the redshift accuracy.  In the HDF,
the simple ML redshifts matched the excellent results in the
literature
\citep{fernandez-soto02,benitez,fontana00,sawicki97,connolly97}, as
shown in Figure \ref{Fig: hdf}, with redshift dispersions of $\sigma_z
\sim 0.07\,(1+z)$ over the full redshift range $0< z < 6$ and only one
``catastrophic error'' \citep[for a discussion of this object
  see][]{fernandez-soto02}.  The photometric redshifts described in
this Section have not been subjected to a Bayesian prior.  In Section
\ref{iterated method} we demonstrate how the use of such a prior improves 
the derived redshift distribution.

\subsection{Calibration Using CFRS Spectroscopic Redshifts}
\label{cfrs}
The accuracy of photometric redshift estimation can be improved using
a spectroscopic training set \citep[see, e.g.,][]{sloan_pz}.  Taking
the CFRS galaxies with the most secure spectroscopic redshifts
(confidence class 3 and 4, see \citealt{cfrs2} for details), we have a
training set of 321 galaxies between $0 \le z \le 1$ and $18.5 \le
I\ab \le 22.5$ (where the bright limit is imposed by the saturation
limit of the CFDF--PRS).  We use this spectroscopic sample as a novel
kind of calibrator to improve the photometric calibration in our three
fields, thereby producing significantly more accurate redshift
estimation.

\begin{figure}[hbtp]
\epsscale{1.25}
\hspace*{-0.6cm}
\plotone{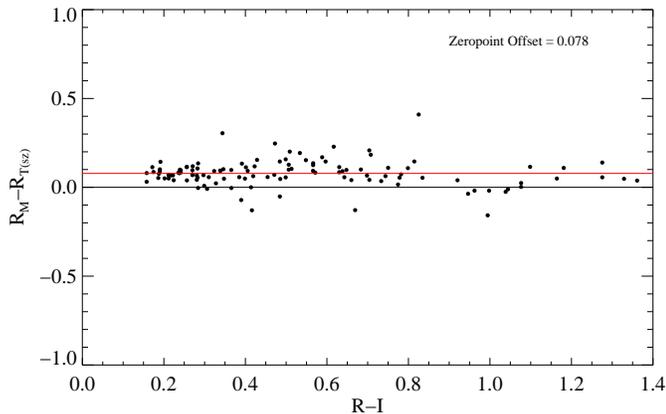}
\caption{Residuals between the measured $R$ magnitudes and the
  best--fitting template SEDs redshifted to the spectroscopic
  redshifts versus galaxy color.  This technique essentially identifies
  the optimal photometric zero point and the effective wavelength in
  each broadband filter, using the CFRS galaxies as photometric
  standards.  In this 03hr $R$ band plot, the zeropoint offset is
  $\sim 0.08$ magnitudes.}
\label{cfrs_calib}
\end{figure}

We examined the residuals between the observed magnitude and the
best--fitting template at the correct (spectroscopic) redshift in each
band, as a function of the color of the galaxies.  As an example, if
all galaxies, regardless of redshift or spectral type, showed a
residual, $\Delta m$, in some band, that would suggest that there was
a zero point offset of $\Delta m$ in that filter.  If $\Delta m$ was
correlated with the color of the objects, that could indicate an error
in the filter effective wavelength (the origin of which is inaccurate
filter transmission or QE curves), or problems with the template SEDs.
In practice such color terms were negligible as expected, whereas
small but significant zeropoint offsets were found in several filters.
These offsets are caused, in part, by band--to--band seeing variations
not fully removed when we convolve our images in each filter to the
same seeing.  Such residual seeing variations would manifest as
zeropoint offsets.  Almost all calculated offsets were less than 0.1
mag, the largest being 0.12 mag.  An example is shown in Figure
\ref{cfrs_calib} for the $R$ filter in the 03hr field.

\begin{figure}[hbtp]
\epsscale{1.25}
\hspace*{-0.6cm}
\plotone{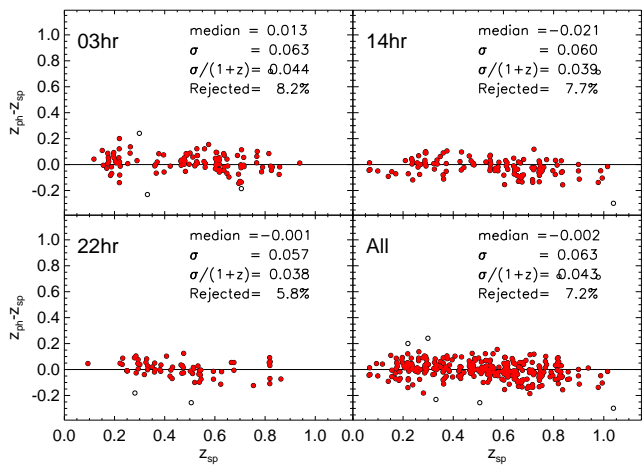}
\caption{Residuals in photometric vs.\ spectroscopic redshift for the
  CFRS spectroscopic galaxies in each field and for the full survey.
  For almost $93\%$ of $I\smallab \leq 22.5$ CFRS galaxies that fall
  within the iteratively clipped 3$\sigma$ limits of the distribution
  (solid circles) the redshift dispersion is $\sigma \sim 0.06$.  In
  the commonly used $\sigma/(1+z)$ measure the dispersion for this
  sample drops to $\sim 0.04$.  Objects represented by open circles
  were rejected as outliers for calculation of the dispersion. }
\label{cfdf_cfrs}
\end{figure}

This optimal calibration produces photometric redshifts with
comparable error properties in all three CFDF--PRS fields, as shown in
Figure \ref{cfdf_cfrs}.  The redshift dispersion, in the common
measure in which the error increases as $(1+z)$ is $\sigma_z/(1+z) =
0.042$ for galaxies within the iteratively clipped $3\,\sigma$ limits
of the distribution (solid circles).  The direct dispersion for this
sample, which includes over $92\%$ of the galaxies is $\sigma_z =
0.062$.

It is important to stress that no bias was introduced to the CFDF--PRS
through the use of the CFRS spectroscopy.  The true redshifts were
used to transform the CFRS galaxies present in the CFDF--PRS into
photometric standards, which were in turn used to tweak our
zeropoints.  Given the spectroscopic redshifts, the process is
completely internal to the CFDF--PRS photometry, and results in
optimal photometric redshifts for our depth and wavelength coverage.

\subsection{Photometric Redshift Accuracy for $I\ab > 22.5$ Galaxies}
\label{Sec: faint sims}
In order to test the accuracy of photometric redshifts at magnitudes
fainter than the CFRS data, we added a scaled version of each image in
each field back to itself, with a positional offset larger than the
full CFRS field.  Care was taken to ensure that the noise properties
of the resulting images were identical to those of the original
images.  Specifically, in each field and filter we multiplied the
original image by a scale factor, $f_1$, translated this scaled image
by $\sim 10\arcmin$, and added it back to the original image.  We then
divided this new image by $(1 + f_1^2)^{0.5}$ to restore the original
image's noise properties.  The scale factor, $f_1$, is chosen such
that in the final image CFRS galaxies with $I\ab =22.5$ are transformed
into $I\ab =24$ objects (i.e.~a 1.5 magnitude change $\rightarrow f_1
\approx 0.2595$).

\begin{figure}[hbtp]
\epsscale{1.25}
\hspace*{-0.6cm}
\plotone{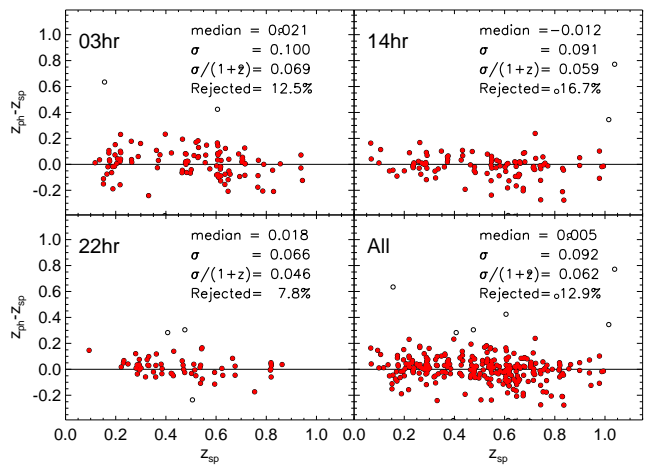}
\caption{Residuals in photometric vs.\ spectroscopic redshift for
  galaxies to $I\smallab \le 24.0$.  The difference in imaging depths
  between the three CFDF--PRS fields is becoming apparent.  In the
  22hr field, the redshift dispersion and number of outliers is
  largely unchanged from the brighter sample in Figure
  \ref{cfdf_cfrs}.  In contrast, the number of outliers in our
  3$\sigma$ clipping algorithm has doubled to 17\% in the 14hr field.
  The redshift dispersions, in each field and overall, are still less
  than $\sigma \sim 0.1$ to $I\smallab \sim 24.0$.}
\label{faintness_sims}
\end{figure}

These fainter CFRS galaxies were photometered, and photometric
redshifts were computed as above.  The results, shown in Figure
\ref{faintness_sims}, demonstrate that while the increasing
photometric error only modestly affects the redshift dispersion in
each field to $I\ab \sim 24$, the fraction of outliers in our
3$\sigma$ clipping algorithm increases to as much as $\sim 15\%$.
Clearly just using the maximum likelihood values of the redshift is
problematic in statistical analyses of these faint galaxies.

While this exercise demonstrates the effect of increased photometric
errors on redshift estimation, it implicitly assumes that $I\ab < 24$
galaxies are basically just fainter copies of CFRS galaxies.  It does
not address the possibility that at this fainter flux limit a new,
distinct galaxy population may emerge, one which may be intrinsically
unlike our empirical galaxy templates.  Photometric redshift accuracy
for such galaxies would likely be quite poor, and the paper's main
results could be substantially compromised.  However, no evidence for
such a hypothetical population is seen in the public HDF, where the
spectroscopic sample extends to $I\ab \sim 25$ and beyond.  Indeed the
excellent photometric redshift accuracy achieved by our code in this
sample indicates that the CWW and \citet{kinney96} templates provide a
good description of galaxies to very faint magnitudes over $0 < z <
6$.  We conclude that the effect on our primary results by any
fundamentally different galaxy population which {\it may} emerge
between $22.5 < I\ab < 24$ is minimal.

\subsection{Photometric Redshift Accuracy for $z \sim 3$ Lyman Break
  Galaxies}
\label{Sec: steidel}

To test the photometric redshift algorithm at redshifts much higher
than the CFRS, we have used the \citet{steidel03} spectroscopic sample
of $z \sim 3$ Lyman Break galaxies (LBGs) present in our 14hr field
(the ``Groth strip'').  We photometer the LBG positions in this field
and compute redshifts for objects with S/N $\ge 3$.  Figure \ref{Fig:
  steidel} shows the redshift residuals for the $104$ spectroscopic
LBGs above this limit.  This sample, being extremely limited in
redshift space, is instructive in exploring the occurrence of
catastrophic redshift errors.  The filled symbols are well measured
redshifts, whereas the open symbols, representing $\sim 13\%$ of the
sample, are catastrophic failures.  Neglecting these the photometric
redshifts at $z \sim 3$ have a dispersion of $\sigma/(1+z) \sim 0.05$.

The catastrophic failures in this LBG sample stem from the difference
in the $U$ filters between the two surveys.  The \citet{steidel96}
bluer $U_n$ filter allows selection of robust LBG candidates at
slightly lower redshifts than our $U$ filter (in which the Lyman
continuum break is only centered at $z \sim 3.2$).  The extremely
faint $z \sim 2.8$ \citet{steidel03} LBGs, without a strong continuum
break fully bracketed by our filters, suffer a degeneracy between the
$z\sim 2.8$ starburst and $0 \la z \la 0.5$ redder galaxy types.
Approximately $20\%$ of the faint $2.7 \le z \le 3.0$ LBGs are
misclassified as lower redshift galaxies.

It is important to note that in this work, in addition to only using
objects with S/N $\ge10$, we are restricting ourselves to a redshift
range which always contains the 4000~\AA\ break.
\begin{figure}[hbtp]
\epsscale{1.25}
\hspace*{-0.6cm}
\plotone{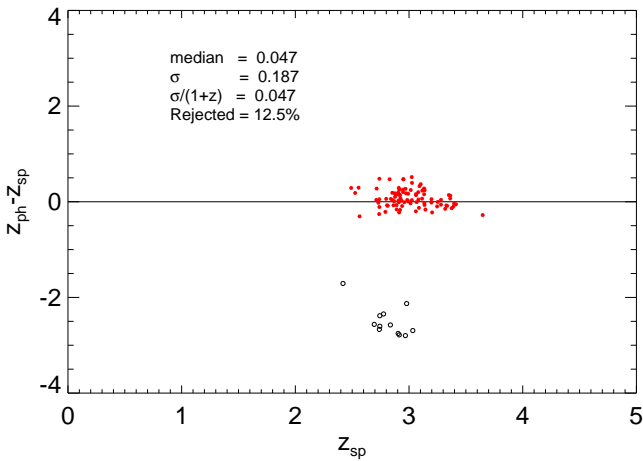}
\caption{Residuals in photometric vs.\ spectroscopic redshift for
  $104$ \protect{\citet{steidel03}} Lyman Break galaxies with S/N $\ge
  3$ in the CFDF--PRS 14hr field.  Excluding catastrophic errors (open
  symbols), the dispersion in redshift is $\sigma/(1+z) \sim 0.05$.
  The galaxies with catastrophic errors, roughly $\sim 13\%$ of the
  sample, are clustered below $z \sim 3$ where our filters do not
  fully straddle a strong continuum break.}
\label{Fig: steidel}
\end{figure}

\subsection{Statistical Validity of Redshift Likelihood Functions:  Confidence Intervals}
\label{Sec: Confidence Intervals}
In Section \textsection{\ref{iterated method}} we introduce a new
method of calculating $N(z)$ in photometric redshift surveys using the
full redshift likelihood function for each galaxy.  Here the error
properties of the likelihood functions are characterized using the
CFRS spectroscopic sample.  Following \citet{fernandez-soto99} we define
the 1, 2, and 3--$\sigma$ confidence intervals as those redshift
regions which enclose the top 68.3\%, 95.4\%, and 99.7\% of the area
in our normalized redshift likelihood functions.  At this point the
likelihood functions have not been subjected to any prior (i.e~there
is equal a priori likelihood at all redshifts).

\begin{figure}[hbtp]
\epsscale{1.25}
\hspace*{-0.6cm}
\plotone{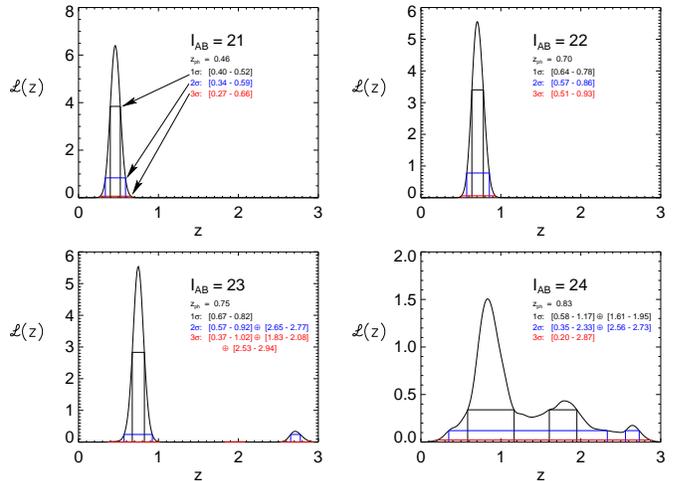}
\caption{Typical likelihood functions with 1$\sigma$, 2$\sigma$, and 3$\sigma$
confidence intervals for increasingly fainter galaxies.  At the survey
limit the likelihood functions become multiply--peaked, leading to
disjoint, but nevertheless statistically valid confidence intervals.}
\label{Fig: likelihoods}
\end{figure}

We found that convolving the likelihood functions with a Gaussian with
$\sigma = 0.04$ produced approximately normal confidence level
statistics, as shown in Table \ref{Table: CL}, and has the beneficial
effect of smoothing the likelihood functions.  We note that the
smoothing kernel is less than our redshift dispersion of $\sim 0.06$
measured with the CFRS spectroscopic sample. Examples of
representative likelihood functions for $I\ab = [21,~22,~23,~24]$
galaxies are shown in Figure \ref{Fig: likelihoods}.

Brighter than $I\ab \la 23.5$ the likelihood functions are
overwhelmingly single--peaked, approximately Gaussian probability
distributions, with relatively little probability density in secondary
peaks.  At the survey limit, however, the secondary peaks become
increasingly significant, often leading to multi--moded, disjoint
confidence intervals even at the 1$\sigma$ level.

\begin{deluxetable}{crr}
\tabletypesize{\normalsize}
\tablecaption{Statistics of confidence intervals derived
from redshift likelihood functions for CFRS spectroscopic galaxies.}
\tablewidth{0pt}
\tablehead{
\colhead{Confidence} & \multicolumn{1}{c}{Observed} & \multicolumn{1}{c}{Expected}\\
\colhead{Interval} & \multicolumn{1}{c}{Number} & \multicolumn{1}{c}{Number}
}
\startdata
$\le 1\sigma$ & 235/321 & 219 \,$^{+16}_{-15}$ \\
$\le 2\sigma$ & 306/321 & 306 \,$^{+19}_{-18}$ \\
$\le 3\sigma$ & 315/321 & 320 \,$^{+19}_{-18}$ \\
$  > 3\sigma$ & 6/321   &  \phn\phn 0.9 \,$^{+2.2}_{-0.8}$ \\
\enddata
\label{Table: CL} 
\end{deluxetable}
The confidence level statistics are nevertheless quite consistent with
a normal error distribution in the Poissonian limit, as shown in Table
\ref{Table: CL}, although we place a slightly higher fraction of
galaxies beyond the 3--$\sigma$ error limit.  The Poisson errors on
the expected number of galaxies are computed using the \citet{poisson}
expressions valid for small numbers of events.  This exercise verifies
the statistical validity of our redshift likelihood functions and of
the analysis presented in Section \textsection{\ref{section: zdist}}
that incorporates them.

\section{Method: Constructing Photometric Redshift Distributions}
\label{iterated method}

For high S/N photometry like that in the HDF, where the redshift
likelihood functions are overwhelmingly narrow and singly--peaked, it
is sensible to simply adopt the redshift which maximizes the
likelihood as the photometric redshift.  However, an open question is
how to best measure photometric redshift distributions in the lower
signal--to--noise photometry typical of ground--based surveys.  The
present method addresses this issue in the CFDF--PRS and in
anticipation of upcoming wide--field near--HDF depth imaging surveys
(such as the CFHT Legacy Survey).

Detailed Monte Carlo simulations (described below) demonstrate that
simple maximum likelihood photometric redshift methods increasingly
fail to reproduce the underlying redshift distribution at S/N
$\lesssim 20$.  This Section describes a Bayesian method for
recovering the underlying $N(z)$ at still more modest S/N levels using
the ensemble redshift information present in the data.

\subsection{Bayesian Photometric Redshift Distributions}

As described in detail in \citet{benitez}, Bayesian statistics provide
a way to incorporate our knowledge about the galaxy distribution to
improve the photometric redshift accuracy.  This is accomplished
through the use of priors on the simple likelihood functions, which by
nature assume all redshifts are a priori equally likely.  The prior
converts the likelihood functions into true Bayesian redshift
probability distributions which can be incorporated directly into
cosmological studies in place of traditional spectroscopic redshifts
(which, in general, can be thought of as approximately
$\delta$--function redshift probabilities).

\noindent Specifically,
\begin{equation}
\label{bayes}
P(z,T) \propto P[z,T|\,N(z)\,] \times {\cal L}(z,T)
\end{equation}
where ${\cal L}(z,T)$ is the likelihood of measuring the observed
colors for a galaxy of spectral type T at redshift z.  In the Bayesian
interpretation $P[z,T|\,N(z)\,]$ is a prior, in this case the
probability of drawing a galaxy with redshift z and type T from a
redshift distribution $N(z)$.  This prior, applied to the likelihood
function, incorporates the basic fact that all redshifts are not a
priori equally likely.  As the goal here is to measure a statistical
property of the galaxy population, namely its redshift distribution,
requiring this information a priori may appear somewhat circular.
However, we have devised an iterative scheme to recover the underlying
redshift distribution and tested it in detailed Monte Carlo
simulations.

\subsection{Direct summation of $N(z)$ and Iteration}
\label{section: iterative method}
Rather than adopting the simple maximum likelihood (ML) value of each
galaxy redshift, we choose instead to directly sum up the normalized
likelihood functions, ${\cal L}^n$, of each galaxy in the
photometric sample.  This would trivially produce the correct redshift
distribution in the limit of spectroscopic (i.e. $\delta$--function)
input redshifts.  For photometric redshifts in the high
signal--to--noise regime this method yields approximately the same
result as collapsing the ${\cal L}^n$ to the single maximum
likelihood redshift since the ${\cal L}^n$ are typically single
sharp Gaussians.  At fainter magnitude limits, the likelihood
functions become broader and often develop multiple peaks as
degeneracies between distinct templates at different redshifts arise
with the increasing photometric error.  With simple ML redshifts the
derived $N(z)$ is both too broad and has artificial peaks as
significant numbers of galaxies are shifted between these degenerate
redshifts.  This is true as well of the first step of our iterative
method:
\begin{equation}
\label{zeroth}
N^0(z) = \sum_i {\cal L}_i^n(z) = \sum_i \frac{{\cal L}_i(z)}{\int_{0}^{\infty} {\cal L}_i(z) \,dz}.
\end{equation}

This zeroth order redshift distribution, $N^0(z)$, is then used as the
prior, $P[z,T|\,N(z)\,]$, in an iterative solution to equation
\ref{bayes}.  This prior uses the bulk information present in the
galaxy colors to modify the raw likelihoods so as to minimize the
degeneracies and, in effect, deconvolve the individual likelihood
functions of the low S/N broadening effect.  The resulting
distribution serves as the first order prior, $N^1(z)$, and the
procedure can be carried on to convergence, which, as shown in the
following Section, is so rapid that additional steps are unnecessary.

This method makes the assumption that a single prior is applicable to
galaxies of all brightnesses and spectral types in a given sample.
Ideally, each population of galaxies should have its own unique prior
to optimally weight the simple likelihood functions.  Although the use
of a single statistical prior, in this case the iteratively estimated
redshift distribution, is obviously a rough approximation, we
demonstrate that it works surprisingly well.  This should not be
mistaken to imply that redshifts of individual galaxies are uniformly
improved with this method.  It is the complete statistical redshift
distribution that is demonstrably improved compared with that obtained
via the standard method of collapsing the redshift likelihood function
of each galaxy to a single maximum likelihood redshift.

Obviously the method breaks down at very low signal--to--noise.
Extensive testing of the convergence properties of the method,
presented in \cite{thesis} and summarized in the Appendix, indicates
that the convergence is excellent over the range $0.2 \le z \le 1.3$
for S/N $\ga 10$.  This is confirmed in the simulations described in
the next Section.

\subsection{Test of the Method:  Monte Carlo Simulation}
\label{montecarlo}
The iterative method described above was extensively tested in
detailed Monte Carlo simulations.  Briefly, model Universes were
populated according to CNOC2 evolving luminosity functions
\citep{cnoc2_lin} for early, intermediate and late spectral types,
modified to produce a range of reasonable input redshift distributions
to $z > 3$.  These model galaxies were assigned colors using the
empirical template set used in the CFDF photometric redshift code,
ranging from the CWW Elliptical \citep{cww} to the \citet{kinney96} SB2.
Flux errors matching the \ubvriz photometry in the CFDF fields were
applied to these galaxies, and photometric redshift likelihood
functions were computed for galaxies in an area of 0.5 deg$^2$.  A
systematic error not accounted for in this simulation is the possible
incompleteness of the adopted template SEDs with respect to the true
range of galaxy properties.  This was approximately accounted for
following the method of \citet{fernandez-soto99} by convolving the
computed redshift likelihood functions with a variable--width
Gaussian,
\begin{equation}
\label{convolution}
{\cal L}^c(z) = \int_{0}^{\infty} dz'\, {\cal L}(z') \,\,G\left[z|z',\Sigma(1+z')\right],
\end{equation}
where G is a normalized Gaussian (truncated at $z<0$) of median z' and
$\sigma=\Sigma(1+z')$.  $\Sigma$ was set to 0.05 based on a detailed
comparison with CFRS spectroscopic redshifts, as described in Section
\textsection{\ref{cfrs}} (in fact, this is $\sim 20\%$ larger than was
found for the CFRS, but was adopted in a conservative vein).

\newpage 
\subsubsection{Simulation Results}

\begin{figure*}
\epsscale{1}
\plotone{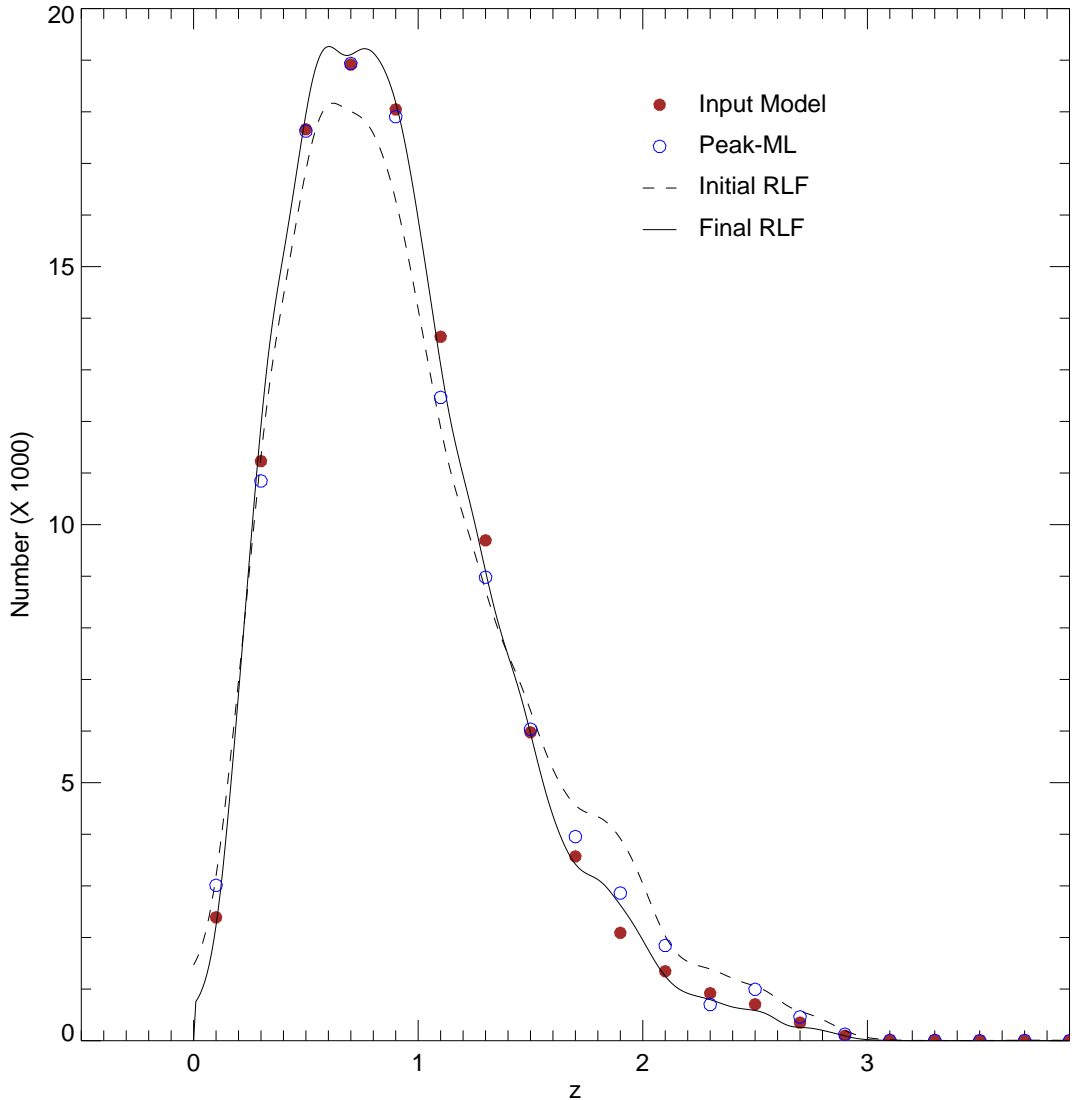}
\caption{Results of Monte Carlo simulations testing Peak--ML (open circles), Initial
  RLF (dashed curve) and Final RLF (solid curve) methods for computing
  $N(z)$ in a photometric redshift survey.  The filled circles show
  the input (model) $N(z)$.  The Peak--ML and Final RLF methods both
  recover the underlying $N(z)$ quite faithfully, but the latter is in
  fact significantly better as discussed in the text and illustrated
  in Figure \ref{dchi^2}.  Note how the aliased peak at $z \sim 1.9$
  in the Initial RLF iteration is strongly suppressed in the Final RLF
  distribution, reproducing the model $N(z)$ better even than the
  Peak--ML method.}
\label{sim_results}
\end{figure*}

\begin{figure*}
\epsscale{1.0}
\plotone{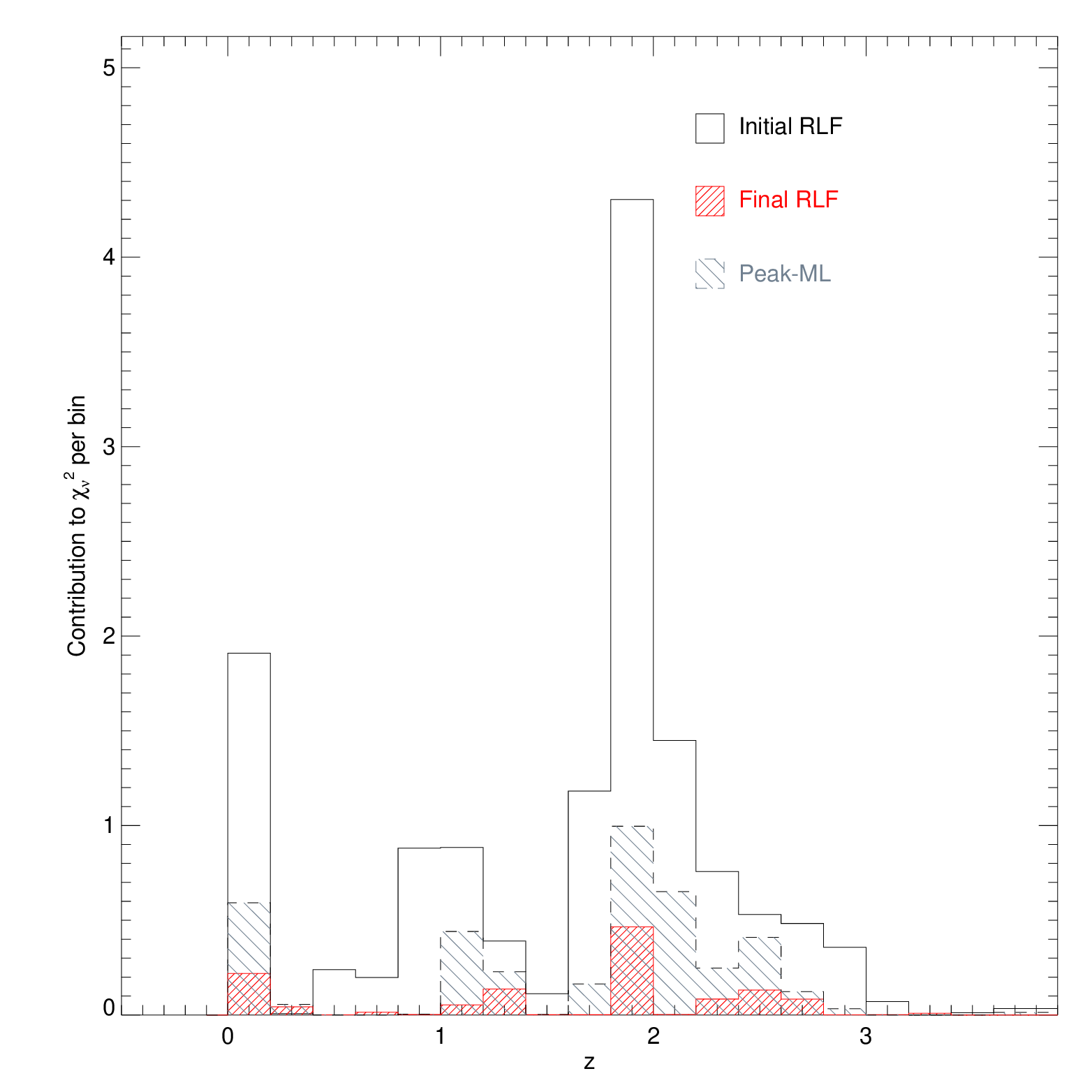}
\caption{Contribution to $\chi_{\nu}^2$ from each redshift bin for the
  Peak--ML (sparsely filled histogram), Initial RLF (empty histogram)
  and Final RLF (densely filled histogram) methods.  Note the
  significant improvement over the Initial RLF method provided by both
  the Peak--ML and Final RLF distributions.  The advantage of the
  Final RLF over the Peak--ML method is also apparent, as the
  differential $\chi_{\nu}^2$ of the former is lower in each bin,
  often dramatically.  The total $\chi_{\nu}^2$ (i.~e.~summing over
  redshift) for these samples is \{Initial RLF, Peak--ML, Final RLF\}
  = \{14.5, 4.02, 1.25\}.}
\label{dchi^2}
\end{figure*}

In the interest of clarity we briefly recall the various methodologies
this simulation was designed to test.  The basic technique for
computing redshift distributions consists of making a simple histogram
of the best--fitting photometric redshifts for each galaxy.  This is
identical to taking the redshift that maximizes the likelihood
function in likelihood-based fitting algorithms.  We therefore dub
this the {\it Peak--ML} method.  Our new method, essentially a
two--step iterative process, uses the full redshift likelihood
function for each galaxy in computing $N(z)$. We therefore label it
the {\it Full--RLF} iterative method (sometimes referred to simply as
the {\it RLF} method for brevity).  Redshift distributions were
computed using both the Peak--ML method and the Full--RLF iterative
method, and compared to the input $N(z)$.  While the final iteration
of the Full--RLF method is superior down to $I\ab \leq 24.0$, the
improvement is most pronounced for the $I\ab \leq 23.5$ sample, which
we plot in Figures \ref{sim_results} and \ref{dchi^2}.

Figure \ref{sim_results} plots the redshift distributions obtained
with the various methods.  The filled and open circles represent the
input (model) and Peak--ML redshift distributions, respectively.  The
dashed and solid lines show the Initial and Final iterations of the
RLF method, respectively.

At first glance, all the methods appear to reproduce the input $N(z)$
quite well.  However, upon closer examination, it is clear that the
Initial RLF $N(z)$ is inferior to both the Peak--ML and the Final RLF
distributions.  On a qualitative level, we see that an incorrect
secondary peak at $z \sim 1.9$ draws galaxies away from the main peak
at $z\sim 0.7$.  This problem is most acute for the Initial RLF
method, but is nevertheless evident in the Peak--ML points.  The Final
RLF method appears to suffer less from the effect of this photometric
redshift degeneracy. 

In fact, the trends observed with the different methods can be easily
understood.  Objects with little redshift ambiguity (i.e.~narrow,
singly--peaked redshift likelihood functions) contribute to the
different $N(z)$ measures in the same, correct way.  On the other
hand, those objects whose redshift likelihood functions show a strong
degeneracy (i.e.~at least two similar peaks at distinct, aliased
redshifts) contribute to $N(z)$ in quite different ways in the above
methods.  In these cases the Peak--ML method usually selects the
correct redshifts, although the catastrophic fraction increases with
decreasing S/N.  The main features of the input $N(z)$ are therefore
fairly well reproduced, although small peaks from the catastrophic
errors will arise at specific aliased redshifts.  The Initial RLF
method, as the direct sum of the likelihood functions, will always
underpredict the main $N(z)$ peak as {\it all} the probability density
of incorrect peaks at aliased redshifts is removed from it, leading to
quite large peaks at the aliased redshifts.  The Final RLF method,
which uses the Initial RLF distribution as a prior, strongly
suppresses these incorrect peaks using the information contained in
the bulk of the galaxy sample (which is, on the whole, correct),
bringing most of the ambiguous redshifts into line.  A quantitative
analysis demonstrates the superiority of the Final RLF method, and
shows that it represents a significant improvement over the Peak--ML
method.

To quantitatively assess the different methods we compare the
resulting $N(z)$ of each method to the input $N(z)$ using a $\chi^2$
test.  To make a fair comparison, we integrate the iterative method
curves over each Peak--ML redshift bin to produce a binned version of
the iterative method for this test.  The Initial RLF redshift
distribution has a reduced $\chi_\nu^2 = 14.5$ when compared with the
input $N(z)$.  The Peak--ML method, unsurprisingly, is much better,
with $\chi_\nu^2 = 4.02$.  It should be remembered that photometric
redshift errors are not formally Gaussian.  In fact in the low S/N
regime, where multiple peaks arise, the error distribution is patently
non--Gaussian, rendering impossible a direct confidence interval
interpretation of the absolute $\chi^2$ values.  However, regardless
of the nature of the photometric error distribution, reductions in
$\chi^2$ should indicate relatively better fits.  In the Final RLF
method the distribution no longer suffers from the effects of the
redshift aliasing.  The curve runs through (or very near) all the
input $N(z)$ data points, and the $\chi_\nu^2$ drops to $1.25$.

Figure \ref{dchi^2} shows the individual contribution from each
redshift bin to the total value of $\chi_\nu^2$ for each method.  The
dramatic improvement from the Initial RLF (empty histogram) to Final
RLF (densely filled histogram) is evident.  The resulting $N(z)$ has
smaller deviations from the input distribution in every bin compared
with the Peak--ML redshift distribution (sparsely filled histogram).

It is interesting to note that in this simulation the iterative method
accurately recovers the redshift distribution outside the
``guaranteed--convergence'' redshift range of $0.2 \lesssim z \lesssim
1.3$ (see Appendix).  Various input redshift distributions and
binnings have been investigated, with results consistent with those
shown in Figure \ref{sim_results}.

In addition to the $\chi^2$ tests Kolmogorov--Smirnov (K--S)
statistics were computed for all methods.  The same pattern emerges,
in which the Peak--ML method produces a redshift distribution more
consistent with the input distribution than the Initial RLF method,
but much less consistent than the Final RLF method.  In this test the
results are not impressive in either of the former two cases, with the
somewhat better Peak--ML redshift distribution matching the input
distribution at a probability of only $10^{-3}$.  This is perhaps not
surprising given the relatively high $\chi^2$ test result.  On the
other hand, the final iteration of the Full--RLF method has a K--S
probability of $0.31$ of matching the input distribution, which is the
level of variation expected in a Poisson sampling of the underlying
distribution.

\section{Redshift Distributions in the CFDF--PRS}
\label{section: zdist}
The iterative method was applied to the CFDF--PRS data described in
Section \textsection{\ref{section cfdf}}.  At bright magnitudes our
results can be compared directly with the CFRS redshift distributions
\citep[hereafter CFRS5]{cfrs5}.

\subsection{Comparison with the CFRS}
\label {Section: cfrs comparison}
The three CFDF--PRS fields contain 3 of the 5 small fields in which
the CFRS survey was carried out.  We can therefore compare the
redshift distributions of each field individually, as well as the
aggregate results for the two surveys to the CFRS limit of $I\ab =
22.5$.  It is important to stress that these surveys are completely
independent.  The imaging data was taken with different cameras and
filters, and the photometric calibrations are entirely independent.
The CFRS magnitudes are isophotal, whereas in the CFDF we use
SExtractor \citep{sextractor} total magnitudes (MAG\_AUTO).

\begin{figure*}
\epsscale{0.8}
\plotone{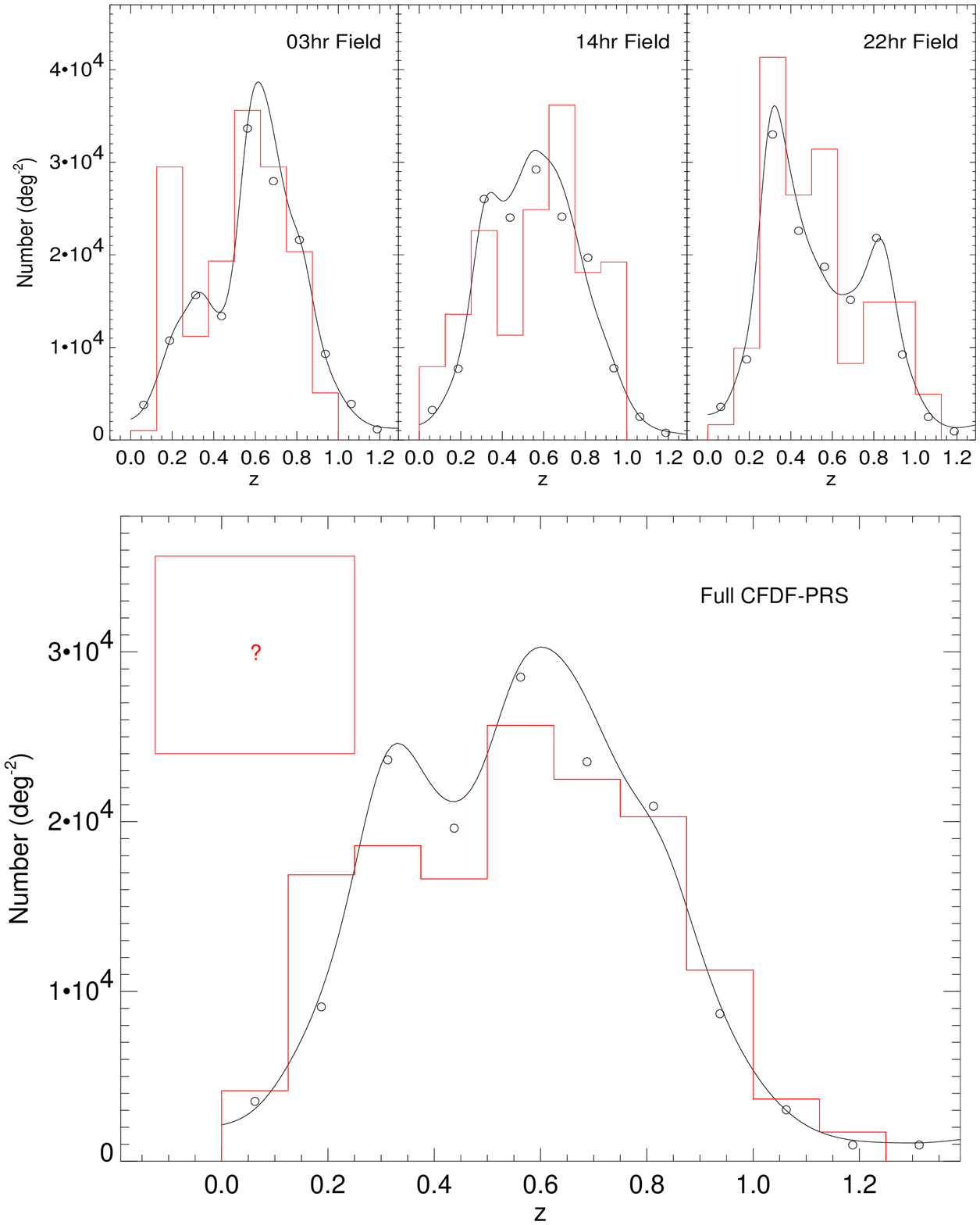}
\caption{CFDF--PRS and CFRS redshift distributions over the common
  magnitude range $18.5 \le I\smallab \le 22.5$.  The curves are the
  Full--RLF final iteration, the open circles represent the Peak--ML
  distributions, and the histograms are from the CFRS.  In the
  individual fields the CFRS number densities are scaled to the CFDF
  values, whereas in the combined fields we plot the CFRS distribution
  using their original normalization.  The fraction of unidentified
  CFRS objects is represented by the box to the left of the peak.
  Stars have been removed from both samples.}
\label{cfdf_cfrs_zdist}
\end{figure*}

Figure \ref{cfdf_cfrs_zdist} compares the $18.5 \le I\ab \le 22.5$
redshift distributions in each common field as well as for the sum of
the fields.  The solid curve is the Final RLF redshift distribution,
the open circles represent the Peak--ML method, and the histogram is
the original CFRS redshift distribution.
The agreement in all three fields is remarkable, though not at all
tautological.  To this depth the CFDF--PRS contains over 20 times the
number of CFRS galaxies in each field.  Nevertheless the striking
field--to--field differences in $N(z)$ discussed in the CFRS persist
in the present much larger survey.  We have conducted tests similar to
those in the CFRS \citep{cfrs5} and conclude that these differences
are not indicative of anisotropies beyond those attributable to
small--scale galaxy correlations.  This impressive agreement between
the CFRS and the CFDF--PRS to $I\ab \le 22.5$ is a firm demonstration
of the excellent control of systematic errors in the survey, as well
as of the accuracy of the photometric redshift code and the iterative
$N(z)$ algorithm.

\subsection{CFDF--PRS $I\ab \le 24$ Redshift Distribution}
\label{Section: I24 zdist}
\begin{figure*}
\epsscale{0.8}
\plotone{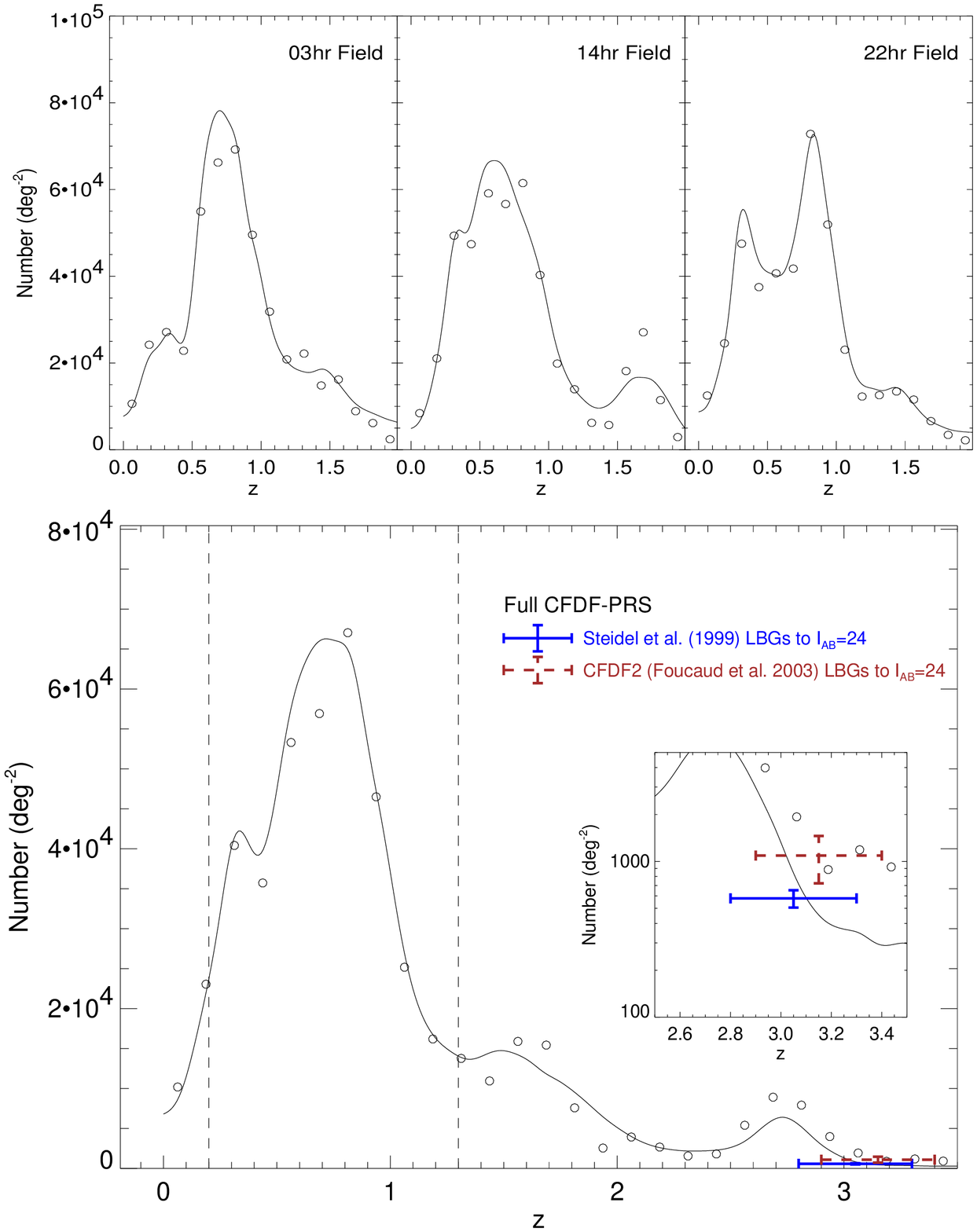}
\caption{CFDF--PRS redshift distribution over magnitude range 
  $18.5 \le I\smallab \le 24.0$. The solid curve and open circles
  represent the Final RLF and Peak--ML redshift distributions,
  respectively.  Also plotted are the \protect{\citet{steidel99}} and
  CFDF2 LBG counts (dashed error bar) integrated to $I\smallab =24$
  for comparison.  The horizontal error bars represent the redshift
  sensitivity ranges of the respective LBG samples, whereas the
  vertical error bars show the 1--$\sigma$ errors in the integrated
  counts.}
\label{cfdf_I24}
\end{figure*}

Pushing beyond the CFRS limit, we present in Figure \ref{cfdf_I24} the
redshift distribution to $I\ab \le 24$ for the full CFDF--PRS.  Once
again the curve is the final iteration of the Full--RLF method, and
the Peak--ML distribution is represented by the open circles.  As
expected, the main difference between these methods in the $0.2 \le z
\le 1.3$ convergence region (between the vertical dashed lines) is a
relative paucity of objects in the Peak--ML method near the median
redshift as these objects were scattered to one of the photometric
redshift degeneracies at $z \sim 1.5$ and $z \sim 2.7$.  The Full--RLF
partially corrects this effect, though a small fraction of objects
remain clustered at these redshifts.

For comparison we plot in the inset the expected $z \sim 3$ Lyman
Break Galaxies to $I\ab \le 24$ from \citet{steidel99}, along with the
number of LBGs from CFDF2 to this depth, both scaled to a one degree
field size.  As explained in Section \textsection{\ref{Sec: steidel}},
the \citet{steidel96} filter set selects slightly lower redshift LBGs
than in the present survey, leading to the small $2.5 \la z \la 3.0$
peak of aliased redshifts.  As we focus on the $0.2 \le z \le 1.3$
convergence region in this work, we refer the reader to CFDF2 for a
more detailed discussion of the low redshift interlopers in the Lyman
Break sample in the CFDF.  We simply note that the LBG densities for
both the \citet{steidel99} and CFDF2 samples lie roughly between the
densities of the Peak--ML and Full--RLF methods, offering a measure of
proof that the present results are reasonable even at $z \gg 1.3$.

The distinct characteristics of each field, as well as the evolution
from the brighter $18.5 \le I\ab \le 22.5$ sample are discussed below.

\subsection{$N(z)$ as a Function of Survey Depth}
\label{Section: field-to-field}
Comparing Figures \ref{cfdf_cfrs_zdist} and \ref{cfdf_I24}, we see
several trends in the transition from the CFRS depth to the CFDF--PRS
limit of $I\ab \sim 24$.  As expected, the $z > 1$ sample grows as a
fraction of the general population, from $\sim 9\%$ to $\sim 30\%$,
while the median redshift increases from $\left<z\right> \sim 0.6$ to
$\left<z\right> \sim 0.8$.  Despite the smoothing effects of
photometric redshift errors, discrete structures in redshift space are
still clearly seen.  In this Section we give a brief description of
the evolution of these structures in each field with survey depth.  It
is interesting to note that these field--to--field differences remain
significant at the full survey depth, even on 30\arcmin\ scales.


\subsubsection{03hr Field {\footnotesize (RA: $03^h03^m00^s$ ~ DEC: $+00\arcdeg09\arcmin00\arcsec$)}}

The 03hr field has considerable structure at $0.6 \la z \la 0.7$,
including a probable cluster near the middle of the field.  This is
confirmed in the CFRS spectroscopic catalog which contains at least
two physical associations of galaxies, each with about 15 galaxies
within $dz \le 0.025$ of systemic redshifts of $z\sim 0.605$ and
$z\sim 0.702$.  Interestingly, the CFRS did not cover the probable
cluster (it being outside their survey area), but its presence
strongly influenced the large scale structure they observed in this
field.  In the CFDF--PRS these two redshift space structures influence
the redshift distribution in this field down to $I\ab = 24$, resulting
in a relative paucity of galaxies at $0.3 \la z \la 0.5$, compared
with the other fields and with the combined redshift distribution.

\subsubsection{14hr Field {\footnotesize (RA: $14^h17^m10^s$ ~ DEC: $+52\arcdeg24\arcmin43\arcsec$)}}

The 14hr field (Groth Strip) is one of the best studied extragalactic
fields \citep[e.g.][]{groth,cfrs3,steidel03}.  It is the most normal,
in some sense, of our three fields.  The picket fence structure
observed in the CFRS has been smoothed out by the photometric redshift
errors, while the thirtyfold increase in area provided by the
CFDF--PRS has reduced the cosmic variance, leaving a relatively
smooth, single--peaked, roughly Gaussian redshift distribution for $z
\le 1.3$.  This distribution does not change appreciably in shape to
$I\ab = 24$, aside from steady increases in median redshift and number
density.  However, at $z\sim 1.8$ we observe a second peak arise in
$N(z)$ as we push to fainter magnitudes.  Visual inspection of the
galaxies involved indicates nothing unusual --- they are entirely
likely to be at the photometric redshift.  However as this peak is
outside our trusted redshift range of $0.2 \le z \le 1.3$ we defer a
detailed study of it until wide--field infrared imaging or
spectroscopic redshifts are available.

The DEEP2 \citep{deep2} redshift survey is targeting this field with
the aim of obtaining 5000 redshifts to $I\ab < 24.5$.  As such, it
will perform a valuable check on the results presented here.

\subsubsection{22hr Field {\footnotesize (RA: $22^h17^m48^s$ ~ DEC: $+00\arcdeg17\arcmin13\arcsec$)}}

The 22hr field is perhaps the most interesting of the three.  There is
a large overdensity at $z\sim 0.3$ observed in both the CFRS and the
CFDF--PRS to the CFRS limit of $I\ab \le 22.5$.  Between $22.5 \le
I\ab \le 24$ the situation changes dramatically, and a broad peak at
$z\sim 0.8$ (the median redshift at this depth) emerges.  The shape of
$N(z)$ in this magnitude interval is quite similar to that in the 03hr
field, albeit at slightly higher redshift.  As the sample selection
deepens from $18.5 \le I\ab \le 22.5$ to $18.5 \le I\ab \le 24$ the
relative sizes of the $z\sim 0.3$ and $z\sim 0.8$ peaks completely
reverse.  In addition, there is evidence for a substantial galaxy
overdensity at $z \sim 1$.  

The VIRMOS--VLT deep spectroscopic survey \cite[VVDS,][]{virmos} will
obtain tens of thousands of spectroscopic redshifts to $I\ab < 22.5$
over 3.6 deg$^2$ in this field.  Given the relatively bright magnitude
limit, it will be instructive to see whether this is sufficient area
to obtain a redshift distribution unbiased by the $z
\sim 0.3$ structure seen in our data, and whether it will be deep
enough to confirm the existence of considerable structure at $z \sim
1$.

\subsection{Median Redshifts in the CFDF--PRS}
\label{Section: Median}

Several statistical studies including the extraction of cosmological
parameters in weak lensing shear studies only require, to good
approximation, the low--order moments of the galaxy population
\citep[e.g.][]{bernardeau97}.  Due to the sensitivity of the mean to
occasional systematic photometric redshift errors, we instead compute
median redshifts in various cumulative and differential magnitude
bins.  Tomographic weak lensing studies can use such information to
improve the measurement of cosmological parameters by up to an order
of magnitude \citep{Hu99}.

While the median is a robust statistic, there are two sources of error
which may nevertheless bias it in a photometric redshift survey such
as CFDF--PRS.  The first, cosmic variance, applies equally well to
traditional spectroscopic surveys.  It is a sort of random error, in
the sense that it can be minimized by observing many fields.
Following \citet{efstathiou91} (see also \citealt{groth_peebles},
\citealt{somerville03}), we calculate the fractional variance in the
counts due to small--scale clustering to be $\sigma^2 \approx 2\%$, or
$\sigma \approx 15\%$.  This agrees quite well with the maximum
empirically measured field--to--field variations given in Tables
\ref{Table: medians} and \ref{Table: R medians}, although the
uncertainties in the latter estimates are themselves of order
$\sqrt{2/(n-1)} = 100\%$ for our $n=3$ fields \citep{pen03}.  The
Poisson noise is negligible for samples as large as this and
therefore, in the absence of systematics, the cosmic variance is
expected to dominate the error budget.

The second source of error, catastrophic failures of the photometric
redshift method caused by redshift aliasing in low S/N photometric
data, is systematic in nature.  The simulations of Section
\ref{montecarlo} indicate the redshift regime which is free of such
aliasing ($0.2 \le z \le 1.3$), and we in fact use this convergence
region in fitting our redshift distributions in the next Section.
However, catastrophic errors do take place outside this convergence
region, as is clear from Figure \ref{Fig: steidel}, and we would like
to investigate the effect of these errors on the median of the full
sample.

We make two estimates of the systematic error due to redshift
aliasing.  The first, and most conservative, adopts the worst--case
scenario that {\it all} redshifts outside the convergence region are
untrustworthy, which from Section \textsection{\ref{Sec: steidel}} we
know is not the case.  We compute the median omitting all objects
outside the $0 \le z \le 1.3$ convergence region, using two one--sided
redshift cuts (i.e.~$z \ge 0.2$ and $z \le 1.3$) to estimate the
asymmetric maximal systematic error possible in the survey.  A more
realistic estimate of the systematic error is estimated via
simulations like those described in Section \textsection{\ref{Sec:
    faint sims}}.  The CFDF--PRS $I\ab$ and $R\ab$ median redshifts,
along with all of the error estimates are listed in Tables \ref{Table:
  medians} and \ref{Table: R medians}, respectively.  We also provide
in these Tables median redshifts in differential (half) magnitude
bins.  Given these results, simple apparent magnitudes cuts in future
imaging surveys can be used as proxies for photometric redshift
information in tomographic weak lensing studies.

\begin{deluxetable*}{cccccccccc}
  \tablecolumns{10} \tabletypesize{\footnotesize}
  \tablecaption{$I\smallab$ Median Redshifts in the
    CFDF--PRS\label{Table: medians}} \tablewidth{0pt} \tablehead{
    \colhead{} & \colhead{} & \colhead{} &\multicolumn{2}{c}{} &&
    \multicolumn{4}{c}{Potential Systematic Effects}\\
    \colhead{}  & \colhead{}    &  \colhead{}  &\multicolumn{2}{c}{Random Errors} && \multicolumn{4}{c}{due to Redshift Aliasing}\\
    \cline{4-5} \cline{7-10} \\
    \colhead{}    &\colhead{}&\colhead{}    & \colhead{1--$\sigma$} & \colhead{1--$\sigma$}   && \multicolumn{1}{c}{$\Delta_{\mbox{best}}$}& \multicolumn{1}{c}{$\Delta_{\mbox{best}}$} & \multicolumn{1}{c}{$\Delta_{\mbox{max}}$} & \multicolumn{1}{c}{$\Delta_{\mbox{max}}$} \\
    \colhead{Mag Range} & \colhead{$\langle I\smallab  \rangle$} &\colhead{$\langle z \rangle$} &
    \colhead{Bootstrap} & \colhead{Field--to--Field}& &
    \multicolumn{1}{c}{($z=3 \rightarrow 0$)} &
    \multicolumn{1}{c}{($z=0 \rightarrow 3$)} &
    \multicolumn{1}{c}{($z \le 1.3$)}& \multicolumn{1}{c}{($z \ge
      0.2$)} } \startdata
  $18.5 \le I\smallab \le 20.5$ &  $19.88$ &  $0.38$ & $0.006$ & $0.035$ && $-0.00$ & $ 0.00$  & $-0.01$ & $0.02$ \\
  $18.5 \le I\smallab \le 21.0$ &  $20.36$ &  $0.43$ & $0.006$ & $0.065$ && $-0.00$ & $ 0.01$  & $-0.01$ & $0.03$ \\
  $18.5 \le I\smallab \le 21.5$ &  $20.79$ &  $0.49$ & $0.007$ & $0.075$ && $-0.00$ & $ 0.01$  & $-0.02$ & $0.02$ \\
  $18.5 \le I\smallab \le 22.0$ &  $21.23$ &  $0.54$ & $0.006$ & $0.070$ && $-0.00$ & $ 0.01$  & $-0.01$ & $0.02$ \\
  $18.5 \le I\smallab \le 22.5$ &  $21.68$ &  $0.59$ & $0.004$ & $0.055$ && $-0.00$ & $ 0.01$  & $-0.02$ & $0.02$ \\
  $18.5 \le I\smallab \le 23.0$ &  $22.11$ &  $0.64$ & $0.004$ & $0.031$ && $-0.00$ & $ 0.01$  & $-0.03$ & $0.02$  \\
  $18.5 \le I\smallab \le 23.5$ &  $22.59$ &  $0.71$ & $0.005$ & $0.040$ && $-0.01$ & $ 0.01$  & $-0.06$ & $0.01$  \\
  $18.5 \le I\smallab \le 24.0$ &  $23.11$ &  $0.77$ & $0.005$ & $0.049$ && $-0.01$ & $ 0.02$  & $-0.09$ & $0.02$  \\
  \\
  \tableline
  \\
  $18.5 \le I\smallab \le 19.0$ &  $18.76$&  $0.30$ & $0.010$ & $0.023$ && $ -0.00$ & $ 0.01$  & $-0.01$ & $0.03$ \\
  $19.0 \le I\smallab \le 19.5$ &  $19.27$&  $0.32$ & $0.008$ & $0.010$ && $ -0.00$ & $ 0.00$  & $-0.01$ & $0.02$ \\
  $19.5 \le I\smallab \le 20.0$ &  $19.79$&  $0.39$ & $0.011$ & $0.023$ && $ -0.00$ & $ 0.01$  & $-0.01$ & $0.02$ \\
  $20.0 \le I\smallab \le 20.5$ &  $20.28$&  $0.43$ & $0.011$ & $0.085$ && $ -0.00$ & $ 0.01$  & $-0.01$ & $0.03$ \\
  $20.5 \le I\smallab \le 21.0$ &  $20.79$&  $0.52$ & $0.010$ & $0.072$ && $ -0.00$ & $ 0.00$  & $-0.01$ & $0.02$ \\
  $21.0 \le I\smallab \le 21.5$ &  $21.28$&  $0.58$ & $0.007$ & $0.050$ && $ -0.00$ & $ 0.01$  & $-0.01$ & $0.01$  \\
  $21.5 \le I\smallab \le 22.0$ &  $21.78$&  $0.63$ & $0.007$ & $0.040$ && $ -0.00$ & $ 0.01$  & $-0.02$ & $0.01$  \\
  $22.0 \le I\smallab \le 22.5$ &  $22.27$&  $0.71$ & $0.006$ & $0.036$ && $ -0.01$ & $ 0.00$  & $-0.03$ & $0.01$  \\
  $22.5 \le I\smallab \le 23.0$ &  $22.77$&  $0.78$ & $0.007$ & $0.031$ && $ -0.00$ & $ 0.01$  & $-0.06$ & $0.02$  \\
  $23.0 \le I\smallab \le 23.5$ &  $23.27$&  $0.89$ & $0.007$ & $0.031$ && $ -0.01$ & $ 0.01$  & $-0.11$ & $0.02$  \\
  $23.5 \le I\smallab \le 24.0$ &  $23.78$&  $1.03$ & $0.008$ & $0.061$ && $ -0.04$ & $ 0.02$  & $-0.23$ & $0.03$  \\
  $24.0 \le I\smallab \le 24.5$\tablenotemark{1} &  $24.27$ & $1.26$ & $0.010$ & $0.083$ && $ -0.13$ & $ 0.04$  & $-0.47$ & $0.06$  \\
  \enddata \tablenotetext{1}{This is formally outside the range
    trusted for accurate photometric redshifts (see text).}
  \tablecomments{Errors of $\pm 0.00$ are smaller than the
    photometric redshift resolution of $0.01$.}
\end{deluxetable*}

\begin{deluxetable*}{cccccccccc}
  \tablecolumns{10} \tabletypesize{\footnotesize}
  \tablecaption{$R\smallab$ Median Redshifts in the
    CFDF--PRS\label{Table: R medians}} \tablewidth{0pt} \tablehead{
    \colhead{} & \colhead{} & \colhead{} &\multicolumn{2}{c}{} &&
    \multicolumn{4}{c}{Potential Systematic Effects}\\
    \colhead{}    & \colhead{}    & \colhead{}  &\multicolumn{2}{c}{Random Errors} && \multicolumn{4}{c}{due to Redshift Aliasing}\\
    \cline{4-5} \cline{7-10} \\
    \colhead{}    &\colhead{}  &\colhead{}    & \colhead{1--$\sigma$} & \colhead{1--$\sigma$}   && \multicolumn{1}{c}{$\Delta_{\mbox{best}}$}& \multicolumn{1}{c}{$\Delta_{\mbox{best}}$} & \multicolumn{1}{c}{$\Delta_{\mbox{max}}$} & \multicolumn{1}{c}{$\Delta_{\mbox{max}}$} \\
    \colhead{Mag Range} & \colhead{$\langle  R\smallab \rangle$} & \colhead{$\langle z \rangle$} &
    \colhead{Bootstrap} & \colhead{Field--to--Field}& &
    \multicolumn{1}{c}{($z=3 \rightarrow 0$)} &
    \multicolumn{1}{c}{($z=0 \rightarrow 3$)} &
    \multicolumn{1}{c}{($z \le 1.3$)}& \multicolumn{1}{c}{($z \ge
      0.2$)} } \startdata
  $18.5 \le R\smallab \le 20.5$ &  $19.79$ &  $0.34$ & $0.006$ & $0.012$ && $ -0.00$ & $ 0.01$  & $-0.01$ & $0.02$ \\
  $18.5 \le R\smallab \le 21.0$ &  $20.26$ &  $0.36$ & $0.005$ & $0.012$ && $ -0.01$ & $ 0.00$  & $-0.01$ & $0.02$ \\
  $18.5 \le R\smallab \le 21.5$ &  $20.71$ &  $0.38$ & $0.005$ & $0.032$ && $ -0.00$ & $ 0.01$  & $-0.01$ & $0.03$ \\
  $18.5 \le R\smallab \le 22.0$ &  $21.16$ &  $0.42$ & $0.005$ & $0.065$ && $ -0.00$ & $ 0.01$  & $-0.01$ & $0.03$ \\
  $18.5 \le R\smallab \le 22.5$ &  $21.65$ &  $0.47$ & $0.005$ & $0.075$ && $ -0.01$ & $ 0.01$  & $-0.02$ & $0.02$ \\
  $18.5 \le R\smallab \le 23.0$ &  $22.12$ &  $0.52$ & $0.005$ & $0.075$ && $ -0.00$ & $ 0.01$  & $-0.02$ & $0.02$ \\
  $18.5 \le R\smallab \le 23.5$ &  $22.59$ &  $0.57$ & $0.004$ & $0.065$ && $ -0.00$ & $ 0.01$  & $-0.02$ & $0.02$ \\
  $18.5 \le R\smallab \le 24.0$ &  $23.09$ &  $0.63$ & $0.005$ & $0.042$ && $ -0.00$ & $ 0.02$  & $-0.04$ & $0.02$ \\
  $18.5 \le R\smallab \le 24.5$ &  $23.61$ &  $0.71$ & $0.003$ & $0.030$ && $ -0.01$ & $ 0.02$  & $-0.09$ & $0.02$ \\
  \\
  \tableline
  \\
  $18.5 \le R\smallab \le 19.0$ &  $18.78$ &  $0.33$& $0.011$ & $0.040$ && $ -0.00$ & $ 0.01$  & $-0.02$ & $0.03$ \\
  $19.0 \le R\smallab \le 19.5$ &  $19.29$ &  $0.32$& $0.009$ & $0.023$ && $ -0.00$ & $ 0.00$  & $-0.01$ & $0.02$ \\
  $19.5 \le R\smallab \le 20.0$ &  $19.78$ &  $0.33$& $0.009$ & $0.010$ && $ -0.00$ & $ 0.00$  & $-0.01$ & $0.02$ \\
  $20.0 \le R\smallab \le 20.5$ &  $20.29$ &  $0.37$& $0.009$ & $0.021$ && $ -0.00$ & $ 0.00$  & $-0.01$ & $0.02$ \\
  $20.5 \le R\smallab \le 21.0$ &  $20.78$ &  $0.39$& $0.009$ & $0.065$ && $ -0.00$ & $ 0.01$  & $-0.01$ & $0.03$ \\
  $21.0 \le R\smallab \le 21.5$ &  $21.27$ &  $0.44$& $0.009$ & $0.068$ && $ -0.00$ & $ 0.01$  & $-0.00$ & $0.03$  \\
  $21.5 \le R\smallab \le 22.0$ &  $21.78$ &  $0.52$& $0.008$ & $0.085$ && $ -0.00$ & $ 0.01$  & $-0.01$ & $0.01$  \\
  $22.0 \le R\smallab \le 22.5$ &  $22.28$ &  $0.55$& $0.006$ & $0.076$ && $ -0.00$ & $ 0.01$  & $-0.01$ & $0.02$  \\
  $22.5 \le R\smallab \le 23.0$ &  $22.78$ &  $0.64$& $0.006$ & $0.025$ && $ -0.00$ & $ 0.00$  & $-0.02$ & $0.01$  \\
  $23.0 \le R\smallab \le 23.5$ &  $23.28$ &  $0.73$& $0.006$ & $0.023$ && $ -0.01$ & $ 0.01$  & $-0.05$ & $0.01$  \\
  $23.5 \le R\smallab \le 24.0$ &  $23.78$ &  $0.85$& $0.007$ & $0.015$ && $ -0.02$ & $ 0.01$  & $-0.12$ & $0.01$  \\
  $24.0 \le R\smallab \le 24.5$ &  $24.28$ &  $1.01$& $0.009$ & $0.040$ && $ -0.04$ & $ 0.02$  & $-0.24$ & $0.04$  \\
  $24.5 \le R\smallab \le 25.0$\tablenotemark{1} & $24.76$  & $1.21$ & $0.010$  & $0.080$ && $ -0.11$ & $ 0.03$  & $-0.42$ & $0.06$\\
  \enddata \tablenotetext{1}{This is formally outside the range
    trusted for accurate photometric redshifts (see text).}
\end{deluxetable*}
The complete error on each median redshift consists of the random
error (a quadrature sum of the bootstrap and field--to--field errors)
as well as the best estimate of the asymmetric systematic errors.  For
the $18.5 \le I\ab \le 24$ sample the median redshift is therefore
$\langle z \rangle = 0.77 \pm 0.049 \mbox{ (random) }\,\,
^{+0.02}_{-0.01} \mbox{ (systematic)}$, where we use angle brackets to
denote the median.  We include median redshifts for the $24.0 \le I\ab
\le 24.5$ and $24.5 \le R\ab \le 25.0$ samples which are formally
outside our trusted magnitude range for accurate photometric
redshifts, as can be seen by the sudden rise in the systematic error
for these samples.  Nevertheless, as the median is a reasonably robust
statistic and of significant value to weak lensing studies, we
tentatively include these two medians for use until deeper photometric
redshift surveys become available.

An important point to realize in interpreting Tables \ref{Table:
  medians} and \ref{Table: R medians} is that the errors on the
medians would persist in an ideal spectroscopic survey.  The random
error estimates are not specific to the photometric redshift method
and would be unchanged for a similar spectroscopic sample.  However,
the spectroscopic redshift samples published to date do not approach
the combination of field size, area and depth of the present survey,
and have therefore been more affected by cosmic variance.  The
best--estimate systematic errors would remain non--zero for the
faintest magnitude limits in a spectroscopic survey, as single line
redshift identifications are inherently problematic and lead to
catastrophic failures for as many as $\sim 10\%$ of the galaxies
\citep[e.g.~see][]{fernandez-soto02}.  An additional systematic
uncertainty, unique to spectroscopic surveys and quite difficult to
estimate, is how spectroscopic incompleteness, most severe for
galaxies with weak or absent emission lines, affects the measured
$N(z)$.

As is clear from the Table, the field--to--field variance dominates
the error budget in all cumulative (and most differential) magnitude
ranges, and as such, the accuracy of the median redshifts in the
CFDF--PRS is no worse than we had undertaken a more costly
spectroscopic survey.  In fact, for the reasons mentioned above it may
be superior.

\subsection{Comparison to COMBO--17, CNOC2 and HDF Median Redshifts}

\begin{figure}
\epsscale{1}
\plotone{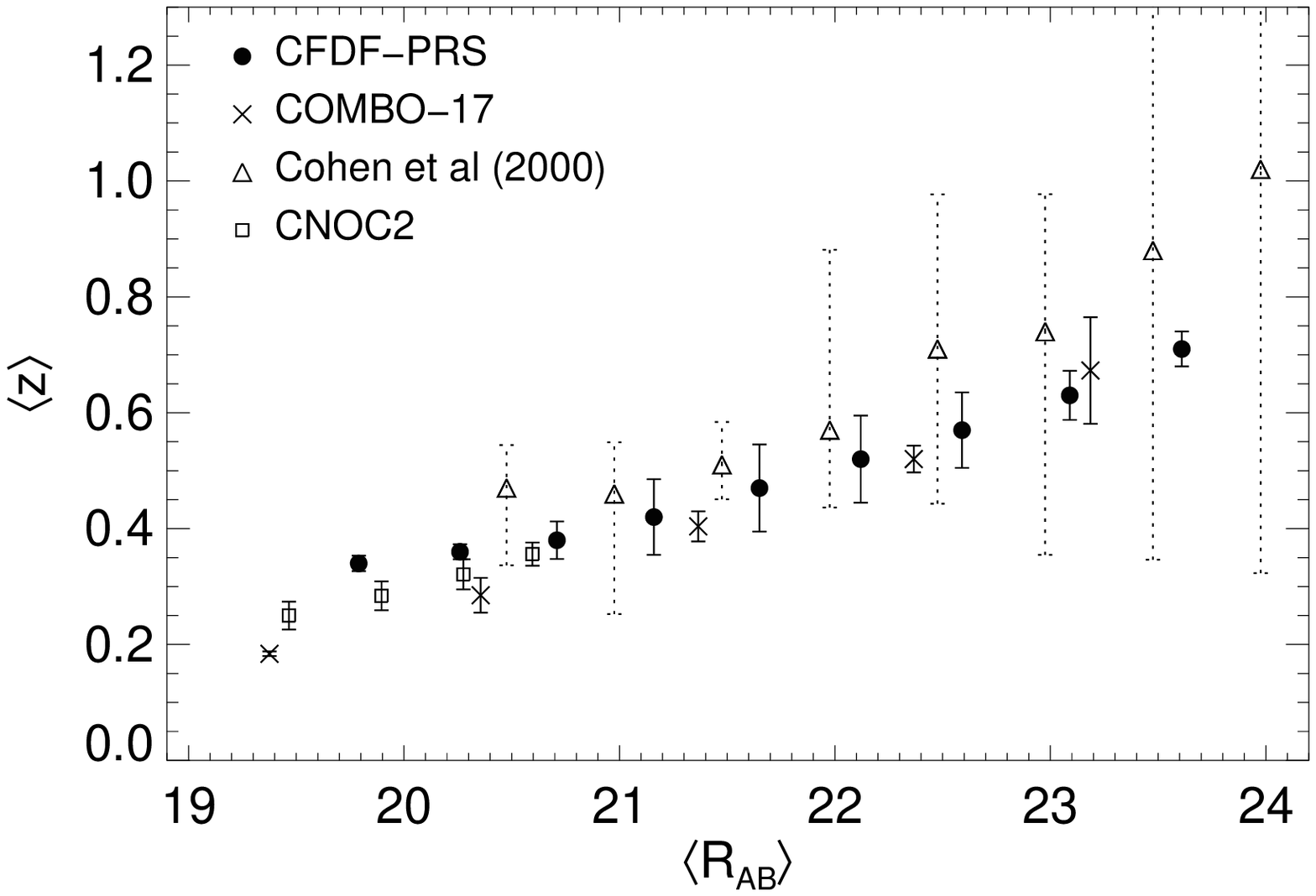}
\caption{Median redshift vs.\ median $R\smallab$ magnitude in the
  CFDF--PRS (filled symbols), along with results for the COMBO--17
  \protect{\citep{combo17_shear}}, CNOC2 \protect{\citep{cnoc2_yee}}
  and HDF North \protect{\citep{cohen00}} surveys , adjusted to AB
  magnitudes.}
\label{Fig: comp_median}
\end{figure}

In Figure \ref{Fig: comp_median} we compare our $R$--band median
redshifts to those of the COMBO--17 \citep{combo17_shear} photometric
redshift survey and the CNOC2 {\citep{cnoc2_yee}} and \citet{cohen00}
spectroscopic redshift surveys.  The median redshifts are plotted
versus median $R\ab$ magnitudes which are tabulated alongside the
various sample magnitude ranges in Table \ref{Table: R medians}.

Our results agree fairly well with, and are in fact intermediate
between, the Combo--17 and \citet{cohen00} results at the bright end,
where the effect of cosmic variance is most important.  For the CNOC2
comparison we compute the median $R\smallab$ magnitudes and redshifts
from the published catalog for the 0223+00 patch, as well as for three
unpublished patches provided to us by H. Yee.  To better match the
CFDF--PRS, we have limited the CNOC2 sample to $R\smallab \ge 18.5$,
used the same magnitude ranges as in the present work, and estimated
the error in the median from the field--to--field variance.  Between
$\langle R\smallab \rangle \sim 20$ and $\langle R\smallab \rangle
\sim 20.75$, the CNOC2 limit, the agreement is excellent.  The
brightest CNOC2 point, at $\langle R\smallab \rangle> \sim 19.5$ is at
somewhat lower redshift than the trend implied by our data, but within
the expected cosmic variance.

At fainter magnitudes these effects are minimized and the comparison
becomes more straightforward.  The agreement with both the COMBO--17
and \citet{cohen00} surveys is excellent for all $R\ab \ga 21$ within
the quoted errors.  For consistency, we transform the published
quartile limits for the \citet{cohen00} sample into 1--$\sigma$ errors
assuming Gaussian distributions.  While we agree with the
\citet{cohen00} median redshifts within these conservatively large
error limits, our fields clearly show a lower median redshift at faint
magnitudes, quite in line with the COMBO--17 results.  In fact, given
the strong agreement our results suggest that for ensemble statistical
measurements of the galaxy population, such as the two--point
correlation function, it is perhaps advantageous to employ only $\sim
6$ broad--band filters (with spectral coverage from $U$ through $Z$
bands) and image more deeply rather than adopt a brighter medium band
approach with $\sim 15$ filters.  While the photometric redshift error
on individual galaxies is likely lower with the medium band survey
strategy, broad--band photometric redshift surveys like the CFDF--PRS
probe deeper down the luminosity function with little penalty on
statistical measurements.

\section{Representation of the CFDF--PRS Redshift Distributions}
\label{Section: fits}

In this Section we present our redshift distributions binned in
magnitude and redshift, as well as in parametrized form in various
magnitude ranges.  It should be noted that none of the main
qualitative results of this paper change if simple maximum likelihood
redshift (Peak--ML) values are used in place of the Final RLF
described in Section \textsection{\ref{iterated method}}.

\subsection{Binned Redshift Distributions}
\label{Section: binned}

In Tables \ref{Table: I binned} and \ref{Table: R binned} we present
the measured CFDF--PRS redshift distributions, binned so as to
present trustworthy information over all redshifts.  Specifically, we
tabulate the Final RLF $N(z)$ in redshift bins of $\Delta z = 0.1$ for
$z < 1.3$ and much more broadly ($\Delta z = 1.0$) at high redshift
where the photometric redshifts are not as accurate.  Our
distributions are presented in $0.5$ magnitude bins, which should
enable other workers to construct redshift distributions for any
sample in the range $18.5 \le I\ab \le 24$, with the obvious caveat
that bright samples containing small numbers of objects have
relatively larger uncertainties than deeper samples.
\begin{deluxetable*}{cccrrrrrrrrrrr}
\tablecolumns{15}
\tabletypesize{\small}
\tablecaption{Binned $I\smallab$ CFDF--PRS Redshift Distributions\label{Table: I binned}}
\tablewidth{0pt}
\tablehead{
& & &\multicolumn{11}{c}{{{\normalsize $\left<I\smallab\right>_{\Delta m = 0.5}$}  }}\\
\\[-0.15cm]
\cline{4-14}
& & &\multicolumn{11}{c}{}\\[-0.1cm]
\colhead{$z_l - z_h$} & \colhead{$z$} && \colhead{$18.75$} & \colhead{$19.25$} & \colhead{$19.75$} & \colhead{$20.25$} & \colhead{$20.75$} & \colhead{$21.25$} & \colhead{$21.75$} & \colhead{$22.25$} & \colhead{$22.75$} & \colhead{$23.25$} & \colhead{$23.75$} }
\startdata
$ 0.05 - 0.15$ & $ 0.10$ && $   157$ & $   161$ & $   140$ & $   266$ & $    313$ & $    284$ & $    343$ & $    419$ & $    636$ & $    928$ & $   1703$ \\
$ 0.15 - 0.25$ & $ 0.20$ && $   320$ & $   390$ & $   362$ & $   533$ & $    726$ & $    750$ & $    956$ & $   1163$ & $   1395$ & $   1648$ & $   2594$ \\
$ 0.25 - 0.35$ & $ 0.30$ && $   638$ & $   905$ & $   869$ & $  1098$ & $   1316$ & $   1485$ & $   1702$ & $   2112$ & $   2080$ & $   2375$ & $   3058$ \\
$ 0.35 - 0.45$ & $ 0.40$ && $   281$ & $   547$ & $   921$ & $  1179$ & $   1404$ & $   1557$ & $   2007$ & $   2147$ & $   2207$ & $   2608$ & $   3255$ \\
$ 0.45 - 0.55$ & $ 0.50$ && $    79$ & $   181$ & $   638$ & $  1042$ & $   1718$ & $   1885$ & $   2787$ & $   2657$ & $   2961$ & $   3224$ & $   4384$ \\
$ 0.55 - 0.65$ & $ 0.60$ && $    48$ & $    78$ & $   387$ & $   968$ & $   2024$ & $   2665$ & $   3551$ & $   3735$ & $   4136$ & $   4544$ & $   5453$ \\
$ 0.65 - 0.75$ & $ 0.70$ && $    27$ & $    29$ & $   153$ & $   401$ & $   1229$ & $   2142$ & $   3355$ & $   4519$ & $   5260$ & $   6073$ & $   6573$ \\
$ 0.75 - 0.85$ & $ 0.80$ && $    10$ & $    10$ & $    44$ & $   162$ & $    562$ & $   1419$ & $   2743$ & $   4264$ & $   5526$ & $   7029$ & $   7506$ \\
$ 0.85 - 0.95$ & $ 0.90$ && $     3$ & $     6$ & $    20$ & $    75$ & $    238$ & $    771$ & $   1617$ & $   2967$ & $   4436$ & $   6511$ & $   7402$ \\
$ 0.95 - 1.05$ & $ 1.00$ && $     1$ & $     3$ & $     9$ & $    22$ & $     66$ & $    224$ & $    659$ & $   1506$ & $   2728$ & $   4899$ & $   6616$ \\
$ 1.05 - 1.15$ & $ 1.10$ && $     1$ & $     2$ & $     4$ & $     7$ & $     19$ & $     62$ & $    237$ & $    670$ & $   1448$ & $   2999$ & $   4870$ \\
$ 1.15 - 1.25$ & $ 1.20$ && $     0$ & $     1$ & $     3$ & $     5$ & $     10$ & $     32$ & $    125$ & $    377$ & $    990$ & $   2119$ & $   3764$ \\
$ 1.25 - 2.25$ & $ 1.75$ && $     0$ & $     1$ & $     2$ & $     2$ & $      5$ & $     16$ & $     62$ & $    199$ & $    512$ & $   1183$ & $   2373$ \\
$ 1.75 - 2.75$ & $ 2.25$ && $     6$ & $     4$ & $     5$ & $     8$ & $      8$ & $     11$ & $     40$ & $     85$ & $    204$ & $    507$ & $   1188$ \\
$ 2.25 - 3.25$ & $ 2.75$ && $    18$ & $     7$ & $     9$ & $    15$ & $     21$ & $     19$ & $     47$ & $     70$ & $    132$ & $    292$ & $    645$ \\
$ 2.75 - 3.75$ & $ 3.25$ && $    14$ & $     3$ & $     5$ & $     8$ & $     13$ & $     10$ & $     16$ & $     24$ & $     48$ & $    116$ & $    289$ \\
$ 3.25 - 4.25$ & $ 3.75$ && $     1$ & $     0$ & $     0$ & $     1$ & $      0$ & $      2$ & $      0$ & $      0$ & $      3$ & $     10$ & $     51$ \\
$ 3.75 - 4.75$ & $ 4.25$ && $     0$ & $     0$ & $     0$ & $     0$ & $      0$ & $      3$ & $      0$ & $      0$ & $      1$ & $      6$ & $     25$ \\
\enddata
\tablecomments{The redshift distributions are presented in constant 0.5 magnitude bins between
$18.5 \le I\ab \le 24$, whereas the $\Delta z = 0.1$ bin size in
redshift is expanded to overlapping $\Delta z = 1.0$ bins at $z >
1.25$.  In each redshift bin the number of galaxies in the $\Delta m =
0.5$ magnitude range is computed as $N = \int_{z_l}^{z_h} N(z, m_l \le
I\ab \le m_h)\, dz /
\int_{z_l}^{z_h} dz$, where $m_l = \left<I\ab\right> - 0.25$, 
$m_h = \left<I\ab\right> + 0.25$, and the number densities, $N$,
represent the CFDF--PRS survey area of 0.4514 deg$^2$.}
\end{deluxetable*}

\begin{deluxetable*}{cccrrrrrrrrrrrr}
\tablecolumns{5}
\tabletypesize{\small}
\tablecaption{Binned $R\smallab$ CFDF--PRS Redshift Distributions\label{Table: R binned}}
\tablewidth{0pt}
\tablehead{
& & & \multicolumn{12}{c}{{{\normalsize $\left<R\smallab\right>_{\Delta m = 0.5}$}  }}\\
\\[-0.15cm]
\cline{4-15} 
& & & \multicolumn{12}{c}{}\\[-0.15cm]
\colhead{$z_l - z_h$} & \colhead{$z$} && \colhead{$18.75$} & \colhead{$19.25$} & \colhead{$19.75$} & \colhead{$20.25$} & \colhead{$20.75$} & \colhead{$21.25$} & \colhead{$21.75$} & \colhead{$22.25$} & \colhead{$22.75$} & \colhead{$23.25$} & \colhead{$23.75$}& \colhead{$24.25$}}
\startdata
$ 0.05 - 0.15$ & $ 0.10$ && $ 128 $ &$ 150 $ & $  183  $ & $  168  $ & $  318 $ & $   279  $ & $   306  $ & $   408  $ & $   429  $ & $   743  $ & $  1051  $ & $  1956  $ \\
$ 0.15 - 0.25$ & $ 0.20$ && $ 230 $ &$ 303 $ & $  432  $ & $  443  $ & $  642 $ & $   782  $ & $   794  $ & $  1171  $ & $  1125  $ & $  1499  $ & $  1722  $ & $  2780  $ \\
$ 0.25 - 0.35$ & $ 0.30$ && $ 488 $ &$ 706 $ & $  850  $ & $ 1017  $ & $ 1243 $ & $  1553  $ & $  1648  $ & $  2088  $ & $  2004  $ & $  2279  $ & $  2475  $ & $  3240  $ \\
$ 0.35 - 0.45$ & $ 0.40$ && $ 340 $ &$ 410 $ & $  605  $ & $  929  $ & $ 1244 $ & $  1658  $ & $  1913  $ & $  2239  $ & $  2224  $ & $  2571  $ & $  2790  $ & $  3410  $ \\
$ 0.45 - 0.55$ & $ 0.50$ && $  92 $ &$ 144 $ & $  222  $ & $  630  $ & $ 1016 $ & $  1802  $ & $  2240  $ & $  3031  $ & $  3096  $ & $  3109  $ & $  3326  $ & $  4204  $ \\
$ 0.55 - 0.65$ & $ 0.60$ && $  43 $ &$  93 $ & $  114  $ & $  302  $ & $  650 $ & $  1587  $ & $  2755  $ & $  3632  $ & $  4382  $ & $  4292  $ & $  4591  $ & $  5037  $ \\
$ 0.65 - 0.75$ & $ 0.70$ && $  11 $ &$  32 $ & $   40  $ & $   63  $ & $  211 $ & $   558  $ & $  1615  $ & $  2618  $ & $  4373  $ & $  5370  $ & $  5893  $ & $  6135  $ \\
$ 0.75 - 0.85$ & $ 0.80$ && $   2 $ &$   8 $ & $    9  $ & $   13  $ & $   43 $ & $    97  $ & $   502  $ & $  1339  $ & $  2882  $ & $  4900  $ & $  6216  $ & $  7061  $ \\
$ 0.85 - 0.95$ & $ 0.90$ && $   1 $ &$   3 $ & $    3  $ & $    3  $ & $   14 $ & $    22  $ & $   133  $ & $   575  $ & $  1624  $ & $  3232  $ & $  5092  $ & $  6597  $ \\
$ 0.95 - 1.05$ & $ 1.00$ && $   0 $ &$   2 $ & $    1  $ & $    2  $ & $    8 $ & $     5  $ & $    40  $ & $   208  $ & $   698  $ & $  1764  $ & $  3377  $ & $  5318  $ \\
$ 1.05 - 1.15$ & $ 1.10$ && $   0 $ &$   1 $ & $    0  $ & $    2  $ & $    4 $ & $     1  $ & $    17  $ & $    80  $ & $   312  $ & $   874  $ & $  2077  $ & $  3737  $ \\
$ 1.15 - 1.25$ & $ 1.20$ && $   0 $ &$   1 $ & $    0  $ & $    2  $ & $    2 $ & $     1  $ & $     9  $ & $    41  $ & $   162  $ & $   537  $ & $  1496  $ & $  2955  $ \\
$ 1.25 - 2.25$ & $ 1.75$ && $   0 $ &$   0 $ & $    1  $ & $    1  $ & $    1 $ & $     1  $ & $     9  $ & $    27  $ & $   101  $ & $   384  $ & $  1031  $ & $  2154  $ \\
$ 1.75 - 2.75$ & $ 2.25$ && $   7 $ &$   3 $ & $    6  $ & $    7  $ & $    8 $ & $     8  $ & $    19  $ & $    41  $ & $    84  $ & $   229  $ & $   565  $ & $  1240  $ \\
$ 2.25 - 3.25$ & $ 2.75$ && $  20 $ &$   8 $ & $   10  $ & $   13  $ & $   20 $ & $    16  $ & $    27  $ & $    45  $ & $    68  $ & $   139  $ & $   297  $ & $   696  $ \\
$ 2.75 - 3.75$ & $ 3.25$ && $  14 $ &$   7 $ & $    5  $ & $    7  $ & $   12 $ & $     9  $ & $    12  $ & $    15  $ & $    27  $ & $    44  $ & $   126  $ & $   341  $ \\
$ 3.25 - 4.25$ & $ 3.75$ && $   1 $ &$   1 $ & $    0  $ & $    1  $ & $    0 $ & $     0  $ & $     1  $ & $     1  $ & $     2  $ & $     2  $ & $    23  $ & $    77  $ \\
$ 3.75 - 4.75$ & $ 4.25$ && $   0 $ &$   0 $ & $    0  $ & $    0  $ & $    0 $ & $     0  $ & $     2  $ & $     1  $ & $     0  $ & $     0  $ & $     3  $ & $    15  $ \\
\enddata
\tablecomments{All quantities are as in defined in Table \ref{Table: I binned}, with
the addition of the distribution for the $24.0 < R\ab < 24.5$ bin in
the final column.}
\end{deluxetable*}

\begin{figure*}
\epsscale{1.0}
\plotone{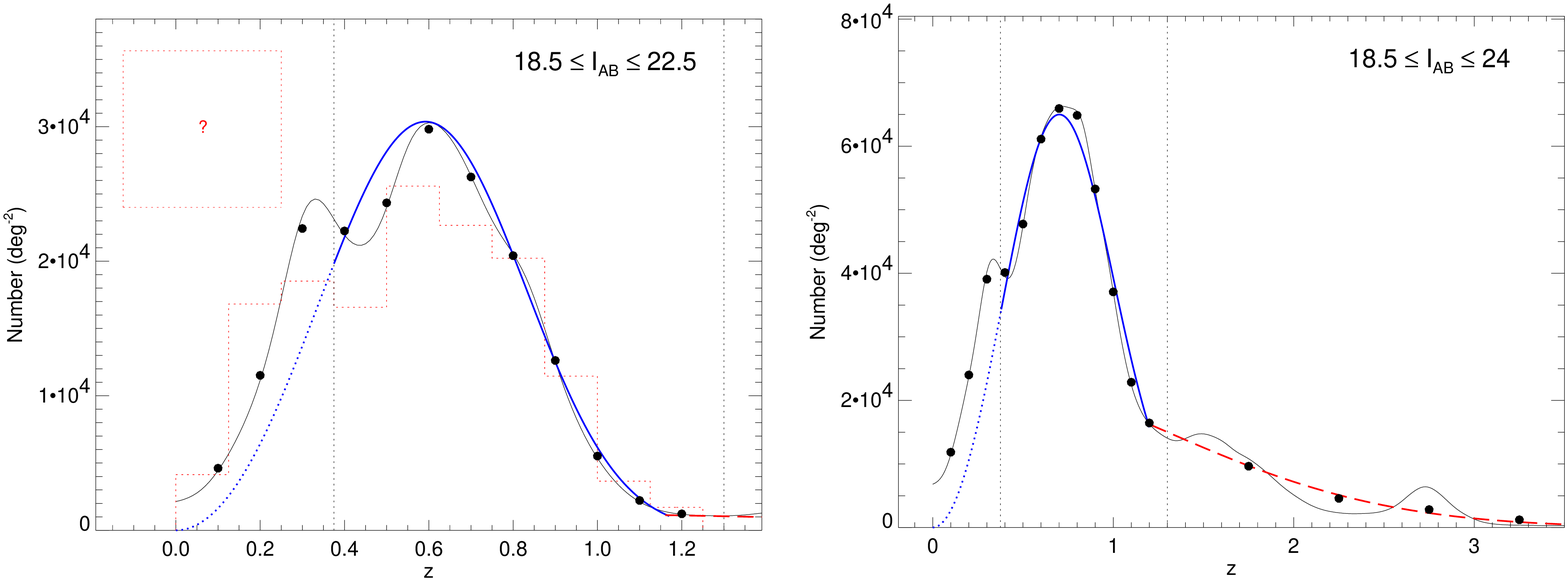}
\caption{Binned distributions (solid symbols) and best--fit
  parametrizations plotted along with the $18.5 \le I~\smallab \le
  22.5$ and $18.5 \le I\smallab \le 24$ RLF redshift distributions
  (thin solid curves), scaled to 1 square degree.  The solid thick
  line shows the best fitting parametrization of the main peak
  (Equation \ref{Eq: model}) at $z \la 1.2$ (represented as a dotted
  curve outside the fitting range, i.~e.~at $z<0.375$).  At $z \ga
  1.2$ the long--dashed curve is the best fitting model of the high
  redshift tail (Equation \ref{Eq: tail}).  The CFRS histogram is
  included for reference as the (dotted) histogram, where once again
  the dotted square represents the unidentified fraction of CFRS
  objects.}
\label{Fig: oplot fits}
\end{figure*}

As an example, in Figure \ref{Fig: oplot fits} we plot the binned
distributions (filled symbols) for the CFRS depth of $18.5 \le I\ab
\le 22.5$ and the full CFDF--PRS depth of $18.5 \le I\ab \le 24$ along
with the full distributions (thin solid curves).  The binned
distributions accurately represent the data at all redshifts, although
they are highly smoothed at $z > 1.25$.  As such our distributions
should be used in this form for precision (e.~g.~cosmic shear)
measurements. 

\subsection{Parametrizations of $N(z)$}

For convenience we provide in this Section parametrized fits to our
redshift distributions, adopting a generalization of the functional
form introduced by \citet{efstathiou91}:
\begin{equation}
N(z)=N_0\frac{\beta}{z_0 \,\Gamma\left(\frac{1+\alpha}{\beta}\right)}
\left(\frac{z}{z_0}\right)^\alpha \exp\left[-\left(\frac{z}{z_0}\right)^\beta\right]
\label{Eq: model}
\end{equation}
These functions were fitted to our $I\ab$ and $R\ab$ redshift
distributions in various magnitude ranges and the results are listed
in Tables \ref{Table: I fits} and \ref{Table: R fits}, respectively.
We limit our fits to the redshift region $0.375 \le z \le 1.3$,
i.~e.~we exclude the $z\sim0.3$ overdensity and fit the main single
peak in each magnitude range.
\begin{deluxetable*}{crrrrrrrrrrr}
\tablecolumns{12}
\tabletypesize{\normalsize}
\tablecaption{Parametrized Fits of $I\smallab$ CFDF--PRS Redshift
Distributions\label{Table: I fits}}
\tablewidth{0pt}
\tablehead{
& &\multicolumn{6}{c}{$z \la 1.2$}&&\multicolumn{3}{c}{$z \ga 1.2$}\\ \\[-0.2cm] 
\cline{3-8} \cline{10-12} 
\\[-0.2cm]
\colhead{Mag Range} && \multicolumn{2}{c}{$N_0$} &
\multicolumn{2}{c}{$z_0$} & \multicolumn{2}{c}{$\beta$} && \colhead{$A_T$} & \colhead{$z_T$} & \colhead{$\gamma$}
}
\startdata
$18.5 \le I\smallab \le 24.0$  && 20785  &  (133) &  \phn \phn 0.790 & (0.008)  &\phn\phn 2.788 & (0.059)  && 11443 & 1.786 & 2.037\\
$19.0 \le I\smallab \le 24.0$  && 20735  &  (141) &  0.793 & (0.009)  & 2.805 & (0.067)  && 11472 & 1.783 & 2.039\\
$19.5 \le I\smallab \le 24.0$  && 20625  &  (143) &  0.800 & (0.008)  & 2.852 & (0.058)  && 11503 & 1.780 & 2.037\\
$20.0 \le I\smallab \le 24.0$  && 20310  &  (135) &  0.818 & (0.009)  & 2.969 & (0.074)  && 11557 & 1.776 & 2.033\\
$20.5 \le I\smallab \le 24.0$  && 19800  &  (126) &  0.840 & (0.008)  & 3.123 & (0.073)  && 11609 & 1.773 & 2.032\\
$21.0 \le I\smallab \le 24.0$  && 18861  &  (141) &  0.872 & (0.007)  & 3.341 & (0.074)  && 11695 & 1.770 & 2.031\\
$21.5 \le I\smallab \le 24.0$  && 17569  &  (128) &  0.902 & (0.007)  & 3.529 & (0.093)  && 11635 & 1.782 & 2.053\\
$22.0 \le I\smallab \le 24.0$  && 15573  &  (120) &  0.945 & (0.008)  & 3.817 & (0.121)  && 11672 & 1.776 & 2.043\\
$22.5 \le I\smallab \le 24.0$  && 13046  &  (111) &  0.978 & (0.009)  & 3.904 & (0.135)  && 11131 & 1.788 & 2.045\\
$23.0 \le I\smallab \le 24.0$  &&  9829  &   (90) &  1.015 & (0.008)  & 3.992 & (0.122)  &&  9383 & 1.837 & 2.074\\
\\ \tableline \\
$18.5 \le I\smallab \le 21.5$  &&   3216 & (169)  &  0.609 & (0.010)  &  3.200 & (0.096) &&     44 & 3.717 & 93.92\\
$18.5 \le I\smallab \le 22.0$  &&   5205 & (165)  &  0.644 & (0.008)  &  3.175 & (0.075) &&    109 & 3.460 & 14.55\\
$18.5 \le I\smallab \le 22.5$  &&   7759 & (186)  &  0.683 & (0.009)  &  3.119 & (0.068) &&    658 & 2.084 & 2.327\\
$18.5 \le I\smallab \le 23.0$  &&  10979 & (168)  &  0.720 & (0.008)  &  3.048 & (0.067) &&   2422 & 1.716 & 1.994\\
$18.5 \le I\smallab \le 23.5$  &&  15293 & (157)  &  0.760 & (0.008)  &  2.941 & (0.066) &&   5889 & 1.699 & 2.013\\
$18.5 \le I\smallab \le 24.0$  &&  20785 & (133)  &  0.790 & (0.008)  &  2.788 & (0.059) &&  11443 & 1.786 & 2.037\\
\enddata \tablecomments{The fits are for the CFDF--PRS survey area of
0.4514 deg$^2$.  At $z \la 1.2$ the best--fitting parameters of
Equation \ref{Eq: model} are listed, along with bootstrap--resampled
errors.  At $z \ga 1.2$, the parameters of fits of the form of
Equation \ref{Eq: tail} are tabulated.  We conservatively estimate the
error in the high redshift tail to be 50\%.}
\end{deluxetable*}

\begin{deluxetable*}{crrrrrrrrrrr}
\tablecolumns{12}
\tabletypesize{\normalsize}
\tablecaption{Parametrized Fits of $R\smallab$ CFDF--PRS Redshift Distributions\label{Table: R fits}}
\tablewidth{0pt}
\tablehead{
& & \multicolumn{6}{c}{$z \la 1.2$}&& \multicolumn{3}{c}{$z \ga 1.2$}\\ \\[-0.2cm] 
\cline{3-8} \cline{10-12} 
\\[-0.2cm]
\colhead{Mag Range} && \multicolumn{2}{c}{$N_0$} &
\multicolumn{2}{c}{$z_0$} & \multicolumn{2}{c}{$\beta$} && \colhead{$A_T$} & \colhead{$z_T$} & \colhead{$\gamma$}
}
\startdata
$18.5 \le R\smallab \le 24.0$  && 13446  &  (133) &  \phn\phn0.662 & (0.012)  & \phn\phn2.601 & (0.081) && 2380 & 2.433 & 3.083\\
$19.0 \le R\smallab \le 24.0$  && 13379  &  (121) &  0.667 & (0.011)  & 2.629 & (0.073) && 2415 & 2.407 & 3.034\\
$19.5 \le R\smallab \le 24.0$  && 13288  &  (113) &  0.673 & (0.010)  & 2.659 & (0.065) && 2425 & 2.399 & 3.025\\
$20.0 \le R\smallab \le 24.0$  && 13151  &  (113) &  0.681 & (0.012)  & 2.710 & (0.081) && 2447 & 2.384 & 2.999\\
$20.5 \le R\smallab \le 24.0$  && 12859  &  (110) &  0.701 & (0.010)  & 2.837 & (0.077) && 2471 & 2.367 & 2.969\\
$21.0 \le R\smallab \le 24.0$  && 12429  &  (122) &  0.725 & (0.010)  & 2.993 & (0.087) && 2507 & 2.343 & 2.927\\
$21.5 \le R\smallab \le 24.0$  && 11659  &  (100) &  0.761 & (0.009)  & 3.232 & (0.091) && 2549 & 2.325 & 2.895\\
$22.0 \le R\smallab \le 24.0$  && 10532  &  (113) &  0.796 & (0.010)  & 3.434 & (0.115) && 2591 & 2.302 & 2.881\\
$22.5 \le R\smallab \le 24.0$  &&  8896  &   (96) &  0.848 & (0.008)  & 3.812 & (0.122) && 2624 & 2.266 & 2.839\\
$23.0 \le R\smallab \le 24.0$  &&  6661  &   (78) &  0.894 & (0.010)  & 4.036 & (0.153) && 2532 & 2.228 & 2.770\\
\\ \tableline \\									 
$18.5 \le R\smallab \le 22.0$  &&   2992 & (170)  &  0.538 & (0.011)  &  3.140 & (0.111)  &&   28 & 3.716 & 76.46\\
$18.5 \le R\smallab \le 22.5$  &&   4607 & (150)  &  0.567 & (0.010)  &  3.116 & (0.084)  &&   70 & 3.670 & 45.30\\
$18.5 \le R\smallab \le 23.0$  &&   6800 & (150)  &  0.602 & (0.011)  &  3.007 & (0.083)  &&  186 & 3.432 & 13.82\\
$18.5 \le R\smallab \le 23.5$  &&   9632 & (145)  &  0.634 & (0.013)  &  2.825 & (0.098)  &&  612 & 2.982 & 5.643\\
$18.5 \le R\smallab \le 24.0$  &&  13446 & (133)  &  0.662 & (0.012)  &  2.601 & (0.081)  && 2380 & 2.433 & 3.083\\
$18.5 \le R\smallab \le 24.5$  &&  18656 & (146)  &  0.665 & (0.012)  &  2.309 & (0.060)  && 5930 & 2.337 & 2.778\\
\tablecomments{All quantities are defined as in Table \ref{Table: I
    fits}.  For the $R\ab$ sample we also provide a fit for the
  $18.5 \le R\ab \le 24.5$ magnitude range.} \enddata
\end{deluxetable*}
Our fitting technique consists of minimizing the Kolmogorov--Smirnov
distance $D$ between the cumulative distribution functions of the
model and Final RLF method redshift distributions over the allowed
redshift range.  The full three parameter fits
[$z_0,\,\alpha,\,\beta]$ have $\alpha$ values very close to the
commonly used value of 2.0 \citep[see, e.g.,][]{efstathiou91},
expected on geometric arguments for a flat average luminosity
function.  Without significantly compromising the accuracy of the
fits, we therefore reduce the dimensionality of the analysis by
setting $\alpha \equiv 2.0$.  The 1--$\sigma$ single parameter errors
in $z_0$ and $\beta$, listed in the Tables, are estimated via 100
bootstrap resamplings (with repetitions) of the CFDF--PRS catalog, and
subsequent recalculation of the redshift distributions using the
Full--RLF method.  The separately determined absolute number
densities, $N_0$, along with bootstrap resampled errors are also
listed in the Tables.

We found Equation \ref{Eq: model} to be a sensible model for $N(z)$ up
to the CFRS depth ($I\ab \sim 22.5$), but at fainter magnitudes it
fails to simultaneously model the main $z\la 1$ peak and the
increasing high redshift tail (at $I\ab \la 24$ the fraction of $z>1$
galaxies is approximately 30\%).  We therefore limit its application
to the $z \la 1.2$ regime and fit the $z \ga 1.2$ tail of the
distribution with a decaying exponential,
\begin{equation}
N(z) = A_T \exp\left[-\left(\frac{z}{z_T}\right)^\gamma\right].
\label{Eq: tail}
\end{equation}

Note that this equation is not fitted to the the Final RLF $N(z)$ but
rather to the binned distributions calculated from Tables
\ref{Table: I binned} and \ref{Table: R binned}.  The transition
between Equations \ref{Eq: model} and \ref{Eq: tail} is always very
close to $z\approx 1.2$, but to produce a continuous parametrization
should in general be taken to be the redshift at which the two
parametrizations intersect.

For the high redshift tail the dominant errors are likely systematic
in nature, coming from the photometric redshift algorithm operating in
a redshift and magnitude regime where the convergence properties are
highly uncertain.  Without a large, deep spectroscopic comparison
sample at $1 < z < 3$ it is difficult to assess these errors in a
meaningful way.  In the present work we conservatively assign a
maximum uncertainty of 50\% to the fit in the $z \ga 1.3$ regime.
Examples of the fits are shown in Figure \ref{Fig: oplot fits} for the
$18.5 \le I\smallab \le 22.5$ and $18.5 \le I\smallab \le 24.0$
samples.  The solid, thick curve is the best fitting parametrization
of the form of Equation \ref{Eq: model} within the $0.375 \le z \le
1.3$ fitting region, although it is likely correct at $z < 0.375$
(dotted curve) due to the constraint at $z=0$.  At $z\ga 1.2$ the
best--fit parametrization of the form of Equation \ref{Eq: tail} is
represented by the long--dashed line.  Taken together the full
redshift range is well parametrized by these functions.

\section{Summary}
\label{summary}

We have introduced a new technique to compute $N(z)$ in which the full
photometric redshift likelihood functions for each galaxy are
incorporated to better reproduce the correct underlying redshift
distribution.  Direct summation of the likelihoods produces the
Bayesian prior which accounts for the fact that all redshifts are not
a priori equally likely.  No information external to the survey is
used in the Bayesian technique, rather we iterate within our own
dataset.  We have presented Monte Carlo simulations which prove the
validity of the technique and demonstrate that it is a significant
improvement over previous methods.

Our highly accurate photometric redshifts, calibrated using hundreds
of spectroscopic CFRS galaxies, have typical dispersions of only
$\sigma/(1+z) \la 0.06$ to $I\ab = 24$ for $z \le 1.3$.  Our large
field sizes ($30\arcmin$) and multiple, widely separated lines of
sight produce redshift distributions far less affected by cosmic
variance than previous surveys of similar depth.  We compute $I\ab$
and $R\ab$ median redshifts, as a function of limiting magnitude and
in differential magnitude bins, with an error budget consisting of
bootstrap resampled random errors, field--to--field variance and an
empirical estimate of the systematic errors due to photometric
redshift aliasing.  We present our $I\ab$ and $R\ab$ redshift
distributions in tabular form and provide parametrized fits with
errors estimated from bootstrap resampling.

This work confirms that in the redshift regimes in which multicolor
imaging surveys have appropriate wavelength coverage and depth, the
errors in photometric redshifts (conservatively $\Delta z/(1+z) \la
0.1$) are not the limiting factor in an accurate determination of
$N(z)$.  In addition, we have verified that the systematic errors
inherent in the method are secondary to cosmic variance in the overall
error budget.  Sparsely sampled spectroscopic or wide-field
photometric redshift surveys containing dozens of effectively
independent fields will be required to significantly improve upon the
present results.

\begin{appendix}
\section{Convergence Properties}

This Appendix presents some of the tests conducted, beyond the Monte
Carlo simulations presented in Section \ref{montecarlo}, to determine
the redshift and S/N ranges for which the new method converges.
\subsection{Formalism}
Generalizing Equation \ref{zeroth} to include a single prior, $P(z)$,
for a given sample
\begin{equation}
{\cal N}(z) =  P(z) \sum_i \frac{{\cal L}_i(z)}{\int_{0}^{\infty} P(z)\, {\cal L}_i(z) \,dz},
\end{equation}
we see that the correct prior is the one for which ${\cal N}(z) = N(z)$
(i.~e.~we recover the correct underlying redshift distribution).  This
leads to the following integral equation for $P(z)$:
\begin{equation}
P(z)=N(z)\left(\sum_i \frac{{\cal L}_i(z)}{\int_{0}^{\infty} P(z)\, {\cal L}_i(z) \,dz}\right)^{-1},
\end{equation}
which can be solved by iteration.  Note that the functional form of
the ideal prior depends on the likelihood functions, as expected since
at very low S/N (reflected in the likelihoods) the method should fail.
Viewing the prior as expressing a change of coordinates to a space in
which the galaxy distribution is flat, one might expect the prior to
be similar to $N(z)$.  Therefore, we can use simulations to determine
the redshift and S/N regimes in which taking $P(z) = N(z)$ produces
the correct answer (i.e.~${\cal N}(z) = N(z)$).  It is
straightforward to show that this requires
\begin{equation}
\label{convergence_ideal}
\eta \equiv \sum_i \frac{{\cal L}_i(z)}{\int_{0}^{\infty} N(z)\, {\cal L}_i(z) \,dz} = 1
\end{equation}
When this equation is satisfied, using $N(z)$ as the prior is strictly
correct.  Although in general $N(z)$ is not known a priori, assuming
that $N^0(z)$ from Equation \ref{zeroth} is a reasonable approximation
to $N(z)$, one might still expect that
\begin{equation}
\label{convergence_real}
\eta^0 \equiv \sum_i \frac{{\cal L}_i(z)}{\int_{0}^{\infty} N^0(z)\, {\cal L}_i(z) \,dz} \approx 1
\end{equation}
is a valid convergence criterion.

\begin{figure}[hbtp]
\epsscale{1}
\plotone{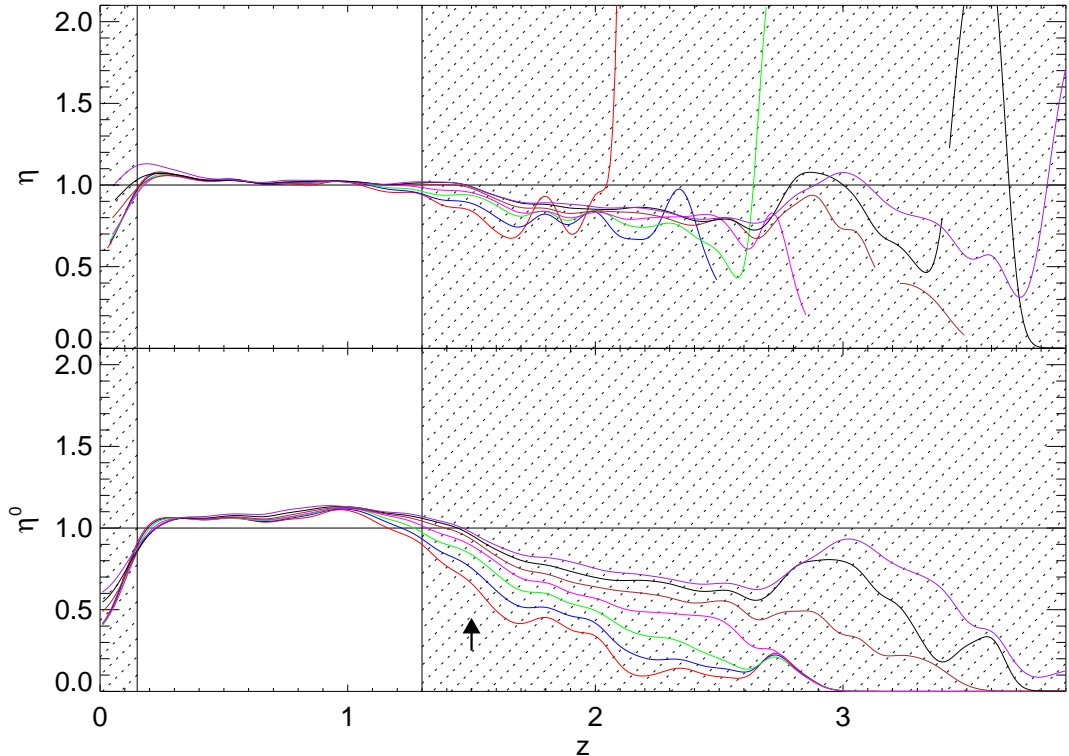}
\caption{Convergence criteria $\eta$ and $\eta^0$ of Equations
  \ref{convergence_ideal} and \ref{convergence_real}, calculated in
  the Monte Carlo simulation of Section \ref{montecarlo}.  The
  unshaded region between $0.2 < z <1.3$ is identified as the most
  trustworthy for recovering the underlying redshift distributions for
  magnitude limited samples down to $I\smallab \sim 24$. Nevertheless,
  the redshift distribution can also be measured outside of this
  redshift range, though slightly less accurately, as shown in the
  next Section.  The curves represent samples down to limiting
  magnitudes of $I\smallab \le 22.5$ (directly above arrow), rising to
  $I\smallab \le 24$ in increments of 0.25 magnitudes.  The curves in
  the top panel are undefined where $N(z)=0$ for bright magnitude
  limits at high redshift.  The smoothing effect of the photometric
  redshift errors prevents this occurrence in the lower panel.}
\label{convergence_graph}
\end{figure}

In the CFDF--PRS we set a conservative $I$--band S/N minimum of 10,
about $I\ab \sim 24$, based on a direct calculation of the convergence
criteria of Equations \ref{convergence_ideal} and
\ref{convergence_real} (shown in Figure \ref{convergence_graph}), as
well as the results of the Monte Carlo simulations described in the
text.  The redshift range over which the method is most reliable to
this depth, apparent from Figure \ref{convergence_graph}, is $0.2
\lesssim z \lesssim 1.3$, corresponding to values of $\eta$ and
$\eta^0$ within 10\%--15\% of unity.  This redshift range is in very
good agreement with the predictions for our filter set
\citep{brodwin99}.  

\end{appendix}
\acknowledgments

We are grateful to Chuck Steidel and Howard Yee for making their
spectroscopic data available to us prior to publication, Chris
Pritchet for observing some of our CFH12k data and Frank Valdes for
assistance with MSCRED.  MB would like to thank Ue--Li Pen, Henk
Hoekstra, Barth Netterfield, and Peder Norberg for useful discussions.
CP was supported by the Zwicky Prize fellowship program at
ETH--Z\"urich.  MB acknowledges support from the National Sciences and
Engineering Research Council and the Ontario Graduate Scholarship
program, as well as ETH--Z\"urich where much of this research was
carried out.

\bibliographystyle{astron2}
\bibliography{/Users/brodwin/Master_bibfile/bibfile.bib}

\end{document}